\documentclass[11pt]{article}
\pdfoutput=1
\usepackage [latin1]{inputenc}
\usepackage{ifpdf}
\usepackage{amsmath}
\usepackage{graphicx}
\usepackage{indentfirst}
\usepackage{amssymb}
\usepackage{cite}
\usepackage{color}
\usepackage{subfigure}
\usepackage[colorlinks,linkcolor=blue,citecolor=blue,urlcolor=blue,hyperindex]{hyperref}
\usepackage{indentfirst}
\usepackage{amsmath}
\usepackage{amssymb}
\usepackage{latexsym}
\usepackage{amsmath}
\usepackage{color}
\usepackage{lineno}
\usepackage{graphicx}
\usepackage{float}
\usepackage{subfigure}
\usepackage{geometry}
\usepackage{graphics}
\usepackage{subfigure}
\usepackage{graphicx}
\usepackage{multirow}
\usepackage{booktabs}
\usepackage{hyperref}
\usepackage{cite}

\begin{document}

\title{Shadows and rings of the Kehagias-Sfetsos black hole surrounded by thin disk accretion}

\date{}
\maketitle

\begin{center}
\author{Guo-Ping Li}$^{a,}$$\footnote{\texttt{Corresponding author: gpliphys@yeah.net}}$ and
\author{Ke-Jian He}$^{b,}$$\footnote{\texttt{kjhe94@163.com}}$

\vskip 0.25in
$^{a}$\it{Physics and Space College, China West Normal University, Nanchong 637000, China}\\
$^{b}$\it{College Of Physics, Chongqing University, Chongqing 401331, China}

\end{center}
\vskip 0.6in
{\abstract
{
In this paper, under the illumination of thin disk accretion, we have employed the ray-tracing method to carefully investigate shadows and rings of the Kehagias-Sfetsos(KS) black hole in deformed Ho\v{r}ava-Lifshitz(HL) gravity.
The results show that the event horizon $r_+$, the radius $r_p$ and impact parameter $b_{p}$ of photon sphere are all decreased with the increase of the HL parameter $\omega$, but the effective potential increases. And, it also turns out that the trajectories of light rays emitted from the north pole direction are defined as the direct emission, lensing ring and photon ring of KS black hole, on the basis of orbits $n = \phi/2\pi$.
As black hole surrounded by thin disk accretion, we show that the corresponding transfer functions have their values increased with the parameter $\omega$. More importantly, we also find that the direct emissions always dominate the total observed intensity, while lensing rings as a thin ring make a very small contribution and photon ring as a extremely narrow ring make a negligible contribution, for all three toy-model functions. In view of this, the results finally imply that shadows and rings as the observational appearance of KS black hole exhibit some obvious interesting features, which might be regarded as an effective way to distinguish black holes in HL gravity from the Schwarzschild black hole.
}
}

\thispagestyle{empty}
\newpage
\setcounter{page}{1}

\section{Introduction}\label{sec1}
Black hole as a prediction of general relativity, has always remained a magic and important object since its concept was proposed.
Until now, it has received many interests both theoretically and experimentally\cite{Hawking1,Hawking2,Hawking3,Hawking4,Hawking5,Hawking6,GW1,Akiyama2019L1,Akiyama2019L2,Akiyama2019L3,Akiyama2019L4,Akiyama2019L5,Akiyama2019L6}.
From theoretical viewpoint, black hole physics has attracted a enormous amount of attention and appeared to bear much fruit over the past few decades\cite{Hawking1,Hawking2,Hawking3,Hawking4,Hawking5,Hawking6}. Also, it is believed that black hole can be regarded as an effective test bed for any proposed schemes for quantum gravity theory. However, it was only recently that some efforts on experiments with respect to black hole have brought out some significant breakthroughs\cite{GW1,Akiyama2019L1,Akiyama2019L2,Akiyama2019L3,Akiyama2019L4,Akiyama2019L5,Akiyama2019L6}.
In 2015, the first discovery of gravitational wave(GW150914) by Laser Interferometer Gravitational wave Observatory (LIGO) provides a direct evidence for the existence of the black hole in our universe\cite{GW1}.
In 2019, the Event Horizon Telescope(EHT) Collaboration has presented the first image of a supermassive black hole in the center of the giant elliptical galaxy M87\cite{Akiyama2019L1,Akiyama2019L2,Akiyama2019L3,Akiyama2019L4,Akiyama2019L5,Akiyama2019L6}, which strongly evidences the general relativity. From this image, it is obvious that there exists a dark central region inside and a bright ring outside, which are now called as black hole shadow and photon ring, respectively.

Black hole shadow, due to the deflection of light, is defined as the dark interior of the critical curve, from which the light ray will asymptotically approach a bound photon orbit. For Schwarzschild black hole, the critical curve occurs at $b_p=3\sqrt{3}M$, and the photon orbit locates at $r_p=3M$, where $M$ is the mass of black hole. At present, shadows together with the related rings as the observational appearance of black holes can give rise to some thoughtful understandings on fundamental properties of black hole, thereby lots of researches related to it have been extensively studied in recent years\cite{GMY1,GMY2,GMY4,wsw1,wsw2,wsw3,wsw4,csb1,csb2,csb3,csb4,csb5,csb6,csb7,csb8,Soo,Shadow2,Shadow3,Shadow4,Shadow5,Shadow6,Shadow7,
Shadow8,Shadow9,Shadow10,Shadow11,Shadow12,Shadow13,Shadow16,Shadow17,Shadow18,Shadow19,Shadow20,Shadow21,Shadow22,add1,add2,add3}.
Bardeen was the first to find that the shape of shadow for Kerr black hole is decided by the rotating parameter, and it can evolve into the D-shape shadow with the increase of rotating parameter\cite{Kerr1,Kerr2}. Then, the shadow casted by a Konoplya-Zhidenko rotating non-Kerr black hole has been discussed and the condition of the D-shape shadow was further presented in reference \cite{csb6}. And, the double shadow of a regular phantom black hole has been studied as photons couple to the Weyl tensor\cite{csb8}. The size and shape of shadow for many other black holes have also been discussed, which are referred to \cite{GMY1,GMY2,GMY4,wsw1,wsw2,wsw3,wsw4,csb1,csb2,csb3,csb4,csb5,csb7,Soo,Shadow2,Shadow3,Shadow4,Shadow5,Shadow6,Shadow7,
Shadow8,Shadow9,Shadow10,Shadow11,Shadow12,Shadow13,Shadow16,Shadow17,Shadow18,Shadow19,Shadow20,Shadow21,Shadow22} and references therein.
However, it is generally believed that the astrophysical black hole in our universe should be surrounded by various kinds of accretion matters.
In view of this, by considering the spherically symmetric accretion, shadows of Schwarzschild black hole have been investigated recently\cite{Gammie}. The result shows that the size of the observed shadow is closely related to the spacetime geometry and there are no influence of the accretion on it\cite{Gammie}. When an optically and geometrically thin-disk accretion surrounded the black hole, the later studies \cite{Wald,ZXX1,ZXX2,GMY3,Li1} find that the black hole image is composed of the dark region and bright region with inclusion of direct emission, lensing rings and photon rings, but the size of it depends on the location of accretion. More importantly, it also turns out that the intensity of observed luminosity for different emitted functions exhibit some interesting features of shadows and rings. Those important features may be regarded as a way for us to distinguish different black holes in the universe. Also, it shows that this method can be used to distinguish the asymmetric thin-shell wormholes from black holes\cite{guoadd}.

On the other hand, due to the absence of quantum gravity theory, the modified gravity theory as an important and hot topic has attracted lots of interest. In those theories, the Ho\v{r}ava-Lifshitz(HL) theory as a renormalizable four-dimensional theory of gravity is always regarded as a good candidate of quantum gravity because it can preserve spatial general covariance and time reparametrization invariance\cite{KS}. And, a great deal of attention is focused on the study of HL gravity. In HL theory, one has found the static spherically symmetric black hole solutions and investigated the related thermodynamic properties\cite{KSb1,KSb2,KSb3,KSt1,KSt2,KSt3}. The quasinormal modes, periodic orbits as well as the gravity lens of those black holes have also been widely discussed\cite{KSg,KSs,KSs1,KSw}. Combined with above facts, our aim in this paper is to carefully investigate the observational appearance of black hole in deformed HL gravity. In particular, when black hole surrounded by the thin disk accretion,
we will employ the ray-tracing method to study the shadows, rings as well as the corresponding observed intensities of luminosity for different emitted functions, and present the differences between our results and that obtained in general relativity.

The organization of the paper is as follows: Sec.\ref{sec2} is devoted to introduce the KS black hole and discuss the effective potential and photon orbits of it. In Sec.\ref{sec3}, when an optically thin-disk accretion surrounded the black hole, we present the shadows, rings as well as the corresponding observed luminosity for a distant observer. Finally, Sec.\ref{Sec4} ends up with conclusions and discussions.

\section{The effective potential and photon orbits of KS black hole}\label{sec2}
In this section, we focus on discussing the effective potential and photon orbits of Kehagias-Sfetsos(KS) black hole.
This black hole solution originated from the HL four-dimensional theory of gravity, and the corresponding action can be expressed as the following form in the limit of cosmological constant $\Lambda_{W} \rightarrow 0 $\cite{KSs,KSg}, which is
\begin{align}\label{eq1}
S_{\rm HL}= \int dt d^3 x (\mathcal{L}_0 + \mathcal{L}_1),
\end{align}
with
\begin{align}\label{eq2}
\mathcal{L}_0 = \sqrt{g} N \left\{ \frac{2}{\kappa^2} \left(K_{i j} K^{i j } - \lambda' K^2 \right) + \frac{\kappa^2 \mu^2 (\Lambda_{W} R - 3 \Lambda_{W}^2)}{8(1-3\lambda')}  \right\},
\end{align}
\begin{align}\label{eq3}
\mathcal{L}_1 =  \sqrt{g} N \left\{ \frac{\kappa^2 \mu^2 (1-4\lambda')}{32(1-3\lambda')} R^2 - \frac{\kappa^2}{2\tilde{\omega}^4} \left(  C_{i j} - \frac{\mu \tilde{\omega}^42}{2} R_{i j} \right)  \left(  C_{i j} - \frac{\mu \tilde{\omega}^42}{2} R^{i j} \right)   + \mu^4 R \right\},
\end{align}
where $\lambda'$, $\mu$, $\kappa$ and $\tilde{\omega}$ are all the parameters of HL gravity, and $K_{i j}$, $C_{i j}$ are the extrinsic curvature and Cotton tensor respectively. Employing the condition $\lambda'=1$ or $\tilde{\omega} = 16\mu^2/\kappa^2$, a static and asymptotically flat black hole solution has been obtained by Kehagias and Sfetsos\cite{KSb3}, it is
\begin{align}\label{eq4}
ds^2=-f (r) d t^2+\frac{1}{f(r)} d r^2+r^2 d \theta ^2 +r^2 {\sin^2\theta}  d \phi ^2,
\end{align}
where $f(r) = 1 + \tilde{\omega} r -\sqrt{\tilde{\omega}^2 r^4 + 4 \tilde{\omega} M r}.$ If one sets $\tilde{\omega} \rightarrow 0 $, the KS black hole will reduce to the Schwarzschild black hole. By rearranging the parameter $\tilde{\omega}$ as
$\omega = \frac{1}{2 \tilde{\omega}^2}$, it is easy to obtain the following form of the metric function,
\begin{align}\label{eq5}
f(r) = 1 + \frac{1}{2\omega^2} \left(r^2 - \sqrt{r^4 + 8 M r \omega^2 }  \right) .
\end{align}
Obviously, the parameter $\omega$ always has a positive value. With the aid of equation $f(r)=0$, the horizons of KS black hole can be obtained, which are
\begin{align}\label{eq6}
r_\pm=M\pm\sqrt{M^2- \omega^2}.
\end{align}
Obviously, $\omega/M \leq 1 $ corresponds to the KS black hole, and $\omega/M =1 $ is related to the extremal case. Combined with above facts, it is obvious that the range of $\omega$ should be fixed to $\omega \in (0,M)$. Next, we shall study the null geodesic of KS black hole at first.
It is well known that, the Euler-Lagrange equation is
\begin{align}\label{eq7}
\frac{d}{d \lambda }\left(\frac{\partial  \mathcal{L}}{\partial \dot{x}^{\mu }}\right)=\frac{\partial \mathcal{L}}{\partial x^{\mu }},
\end{align}
where, $\lambda$ and $\dot{x}^{\mu }$ are the affine parameter and the four-velocity of the light ray, respectively. In this spacetime, the Lagrangian $\mathcal{L}$ for a particle can be written as
\begin{align}\label{eq8}
\mathcal{L}=\frac{1}{2}g_{\mu \nu }\dot{x}^{\mu }\dot{x}^{\nu }=\frac{1}{2}\left[-f(r)\dot{t}^2+\frac{\dot{r}^2}{f(r)}+r^2\left(\dot{\theta }^2+\sin ^2\theta \dot{\phi }^2\right)\right].
\end{align}
In general, we set $\theta = \frac{\pi}{2}$, which implies that the motion of photons is always in the equatorial palne. Meanwhile, since the time $t$ and azimuthal angle $\phi$ are not included in metric function $f(r)$, there are two conserved quantities, which are
\begin{align}\label{eq9}
E = -\frac{\partial \mathcal{L}}{\partial \dot{t}} = f(r)\dot{t} , \qquad L = \frac{\partial \mathcal{L}}{\partial \dot{\phi}} = r^2 \dot{\phi}.
\end{align}
By using Eqs(\ref{eq5}), (\ref{eq7}), (\ref{eq8}) and (\ref{eq9}), and considering the fact $g_{\mu \nu }\dot{x}^{\mu }\dot{x}^{\nu }= 0$ for null geodesic, we have
\begin{align}
&\dot{t}=\frac{1}{b_{c}  \left[ 1 + \frac{1}{2\omega^2} \left(r^2 - \sqrt{r^4 + 8 M r \omega^2 }  \right) \right] } \label{eq100} \\ \quad
&\dot{\phi} = \pm \frac{1}{r^2} \label{eq101}  \\  \quad
&\dot{r}^2= \frac{1}{b_{ c}^2} - \frac{1}{r^2}\left[ 1 + \frac{1}{2\omega^2} \left(r^2 - \sqrt{r^4 + 8 M r \omega^2 }  \right) \right],\label{eq102}
\end{align}
where, the sign $\pm$ in Eq.(\ref{eq101}) represents the counterclockwise and clockwise direction of the light ray. Also, Eq.(\ref{eq102}) can be reexpressed as
\begin{align}\label{eq11}
\dot{r}^2+V_{{ eff}}=\frac{1}{b_{c}^2},
\end{align}
where,
\begin{align}\label{eq12}
V_{{ eff}}=\frac{1}{r^2} \left[ 1 + \frac{1}{2\omega^2} \left(r^2 - \sqrt{r^4 + 8 M r \omega^2 }  \right) \right].
\end{align}
In above equations, $b_c = \frac{|L|}{E}$ is the impact parameter, and $V_{{ eff}}$ is the effective potential of KS black hole, while the affine parameter $\lambda$ has been replaced by $\lambda/|L|$. At the position of the photon sphere, the effective potential should satisfy the conditions,
\begin{align}\label{eq13}
V_{{ eff}}=\frac{1}{b_{ c}^2}, \quad V_{{ eff}}'=0.
\end{align}
For a four-dimensional symmetric black hole, the condition (\ref{eq13}) can be simplified as a relation of the radius $r_{ p}$ and impact parameter $b_{ p}$, which is
\begin{align}\label{eq14}
r_p{}^2=b_{ p}^2 f(r),\quad 2 b_p^2 f(r)^2=r^3 f'(r).
\end{align}
Based on Eq.(\ref{eq14}), the numerical results of the horizon $r_+$, the radius $r_p$ and impact parameters $b_p$ of the photon sphere for different values of parameter $\omega$ have been given in Table 1. It is easy to see that the event horizon $r_+$, the radius $r_p$ and impact parameters $b_p$ are all smaller and smaller with the increase of parameter $\omega$.
\begin{center}
{\footnotesize{\bf Table 1.} The event horizon $r_+$, the radius $r_p$ and impact parameter $b_{p}$ of photon sphere for different values of $\omega$, where $M = 1$\footnote{Note that, the values of $M$ for all figures in the following are all set to $1$ unless are specifically explained. }.\\
\vspace{1mm}
\begin{tabular}{ccccccccccc}
\hline &{$\omega=0.1$} &{$\omega=0.2$} &{$\omega=0.3$} &{$\omega=0.4$} &{$\omega=0.5$} &{$\omega=0.6$} &{$\omega=0.7$} &{$\omega=0.8$} &{$\omega=0.9$} \\ \hline
{$r_{+}$}   &{1.99499}     &{1.97980}      &{1.95394}     &{1.91652}     &{1.86603}     &{1.80000}     &{1.71414}     &{1.60000}     &{1.43589}        \\
{$r_{p}$}   &{2.99555}     &{2.98206}      &{2.95917}     &{2.92619}     &{2.88202}     &{2.82498}     &{2.75242}     &{2.65991}     &{2.53932}        \\
{$b_{p}$}   &{5.19230}     &{5.18064}      &{5.16092}     &{5.13264}     &{5.09505}     &{5.04700}     &{4.98679}     &{4.81705}     &{4.91167}        \\
\hline
\end{tabular}}
\end{center}

\begin{figure} [!h]
\centering
\subfigure[$\omega=0.1$]{
\includegraphics[scale=0.55]{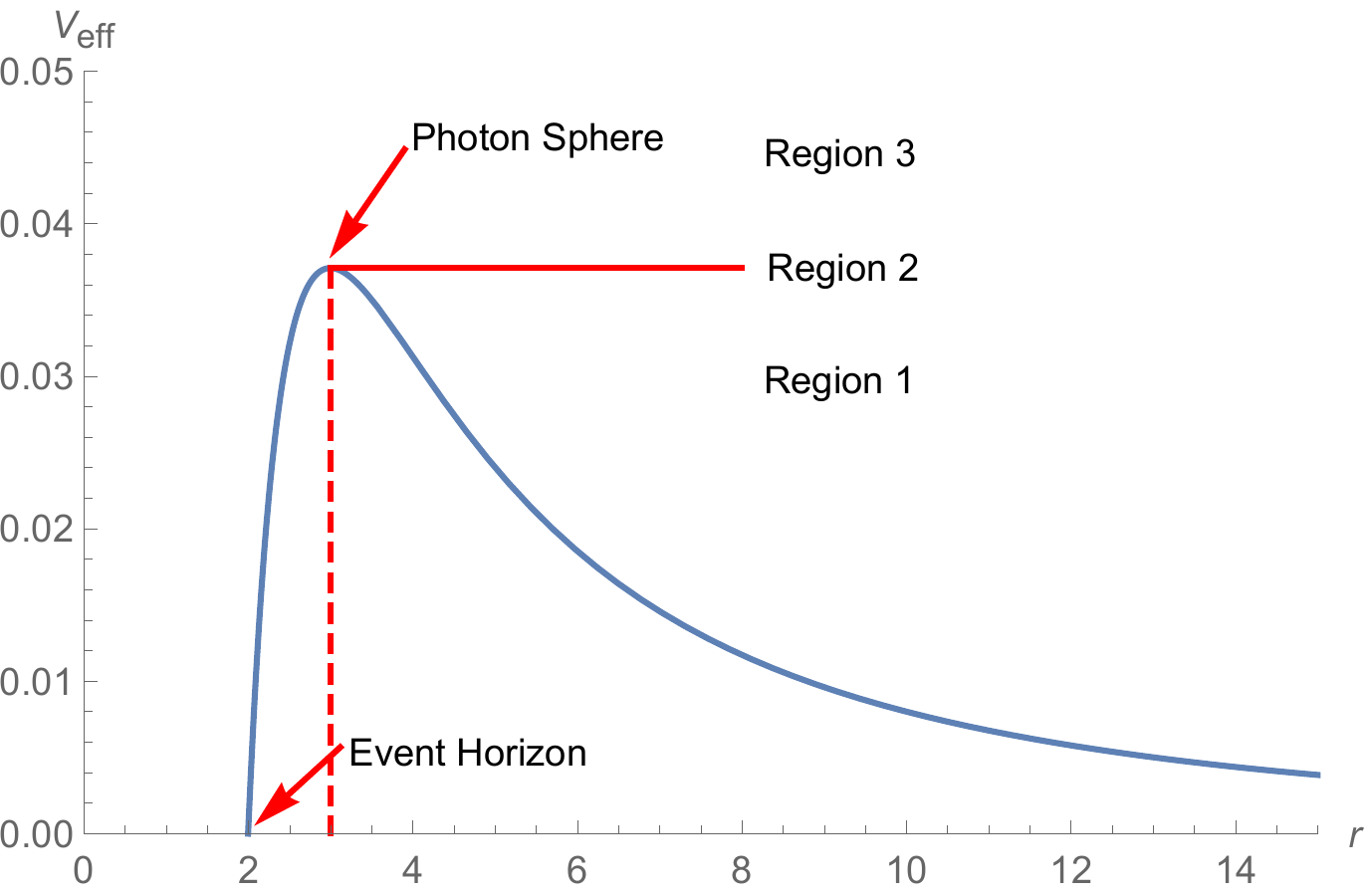}}
\subfigure[$\omega=0.8$]{
\includegraphics[scale=0.55]{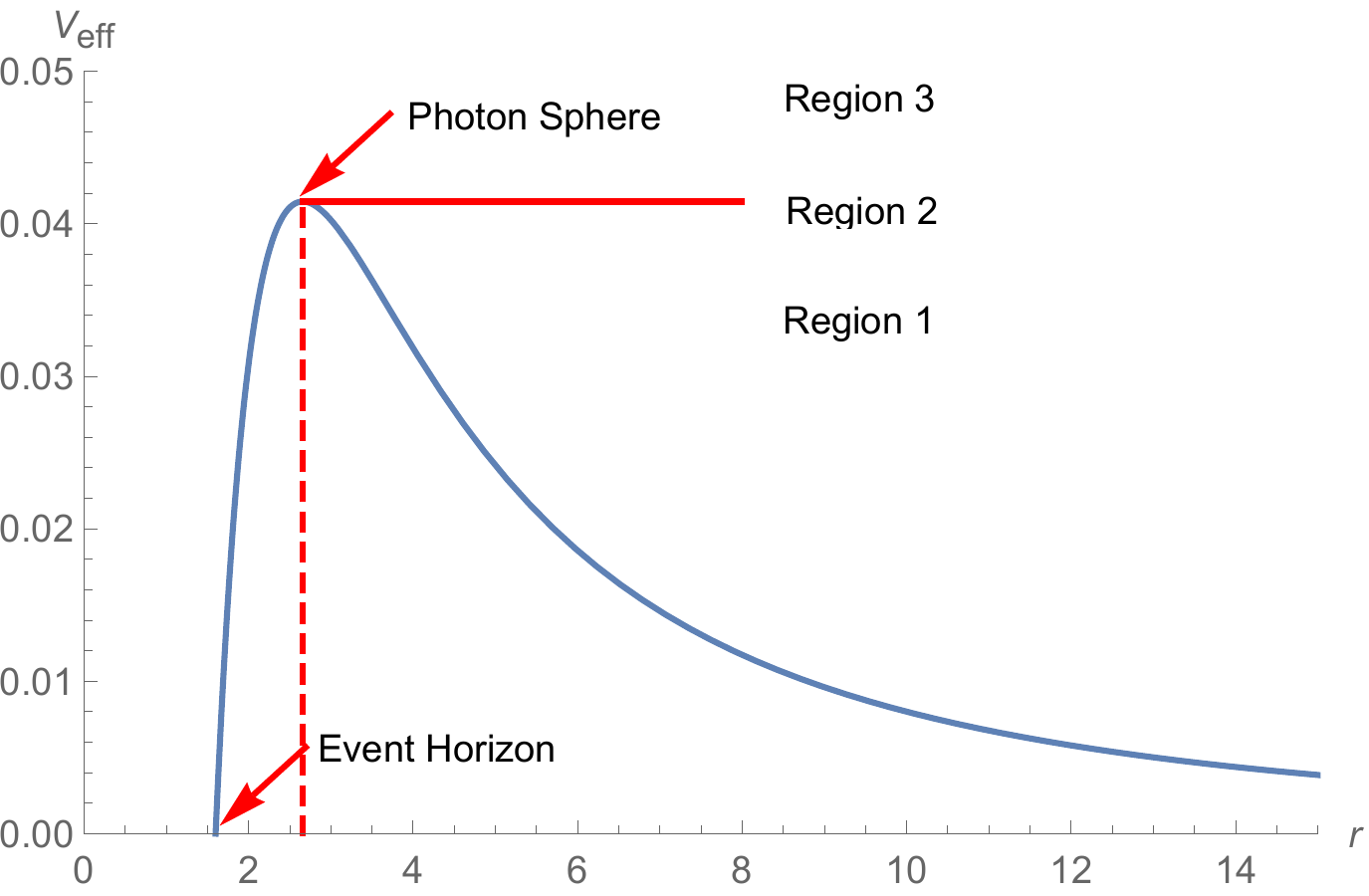}}
\caption{\small  The profiles of the effective potential $V_{eff}$ and impact parameter $b_c$. }\label{figure1}
\end{figure}
In addition, Figure \ref{figure1} shows the effective potential $V_{eff}$(or impact parameter $b_c$) versus the radius $r$ for different values for $\omega$. At the event horizon, one can see that the effective potential $V_{eff}$ has its value be zero. At the photon sphere, there is a maximum value of the effective potential $V_{eff}$, and a larger value of $\omega$ a larger maximum value of $V_{eff}$. And, the effective potential increases with the radius $r$ between this two position, decreases when the radius $r$ is greater than the radius of photon sphere. According to this fact, when a light ray at the infinity moves in the radially inward direction, there are three different Regions in which the light ray has different behaviors of motion. In the Region 1, namely $b_c > b_p$, the light ray will be reflected since it encountered a potential barrier produced by the effective potential $V_{eff}$. In the Region 2, namely $b_c = b_p$, when the impact parameter of the light ray is infinitely close to the radius of photon sphere, the light ray will revolve around the black hole infinitely many times. In the Region 3, namely $b_c < b_p$, because there is no any barriers exist to affect the behavior of the light ray, it will drop into the black hole.

\begin{figure}[!h]
\centering
\subfigure[$\omega=0.1$]{
\includegraphics[scale=0.5]{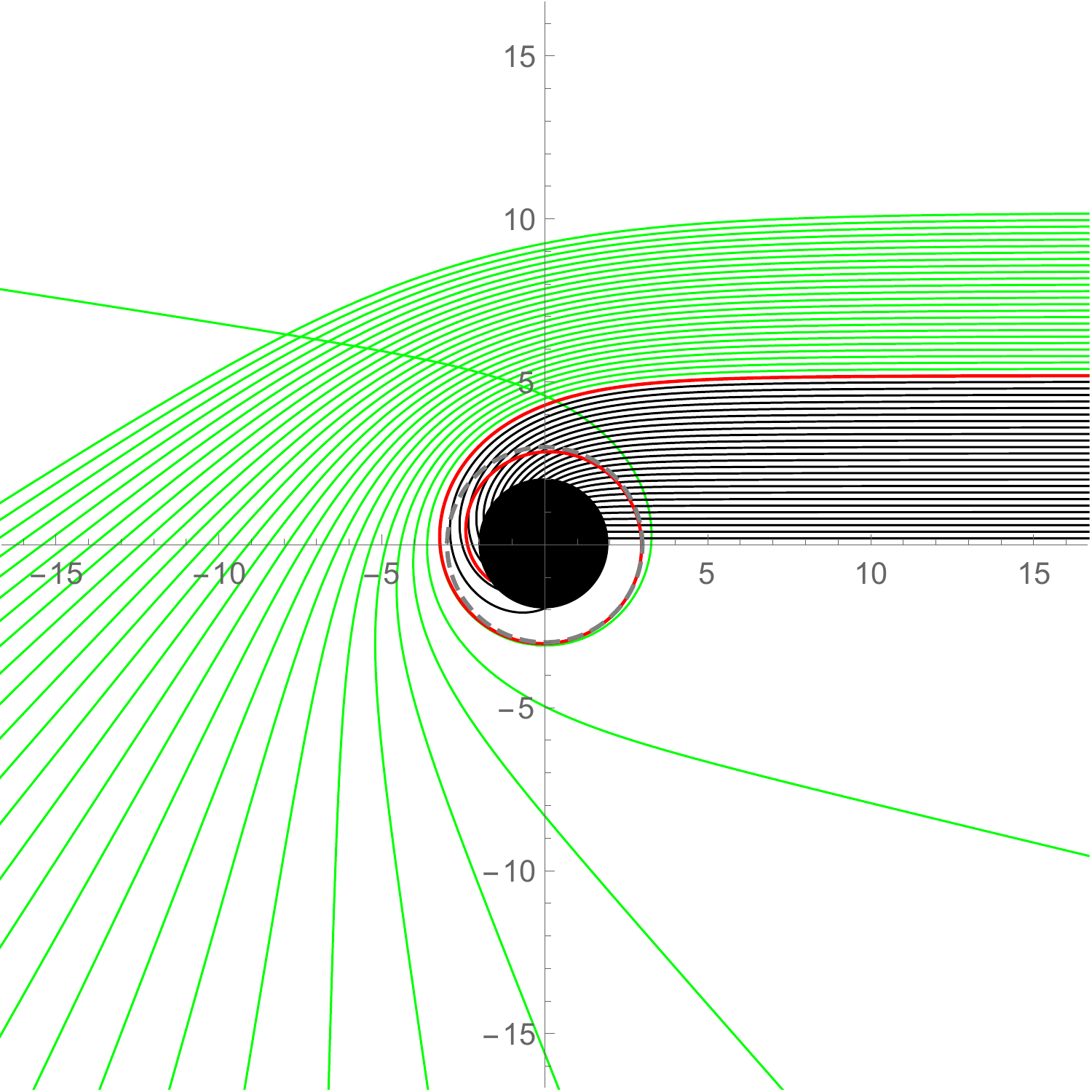}}
\subfigure[$\omega=0.8$]{
\includegraphics[scale=0.5]{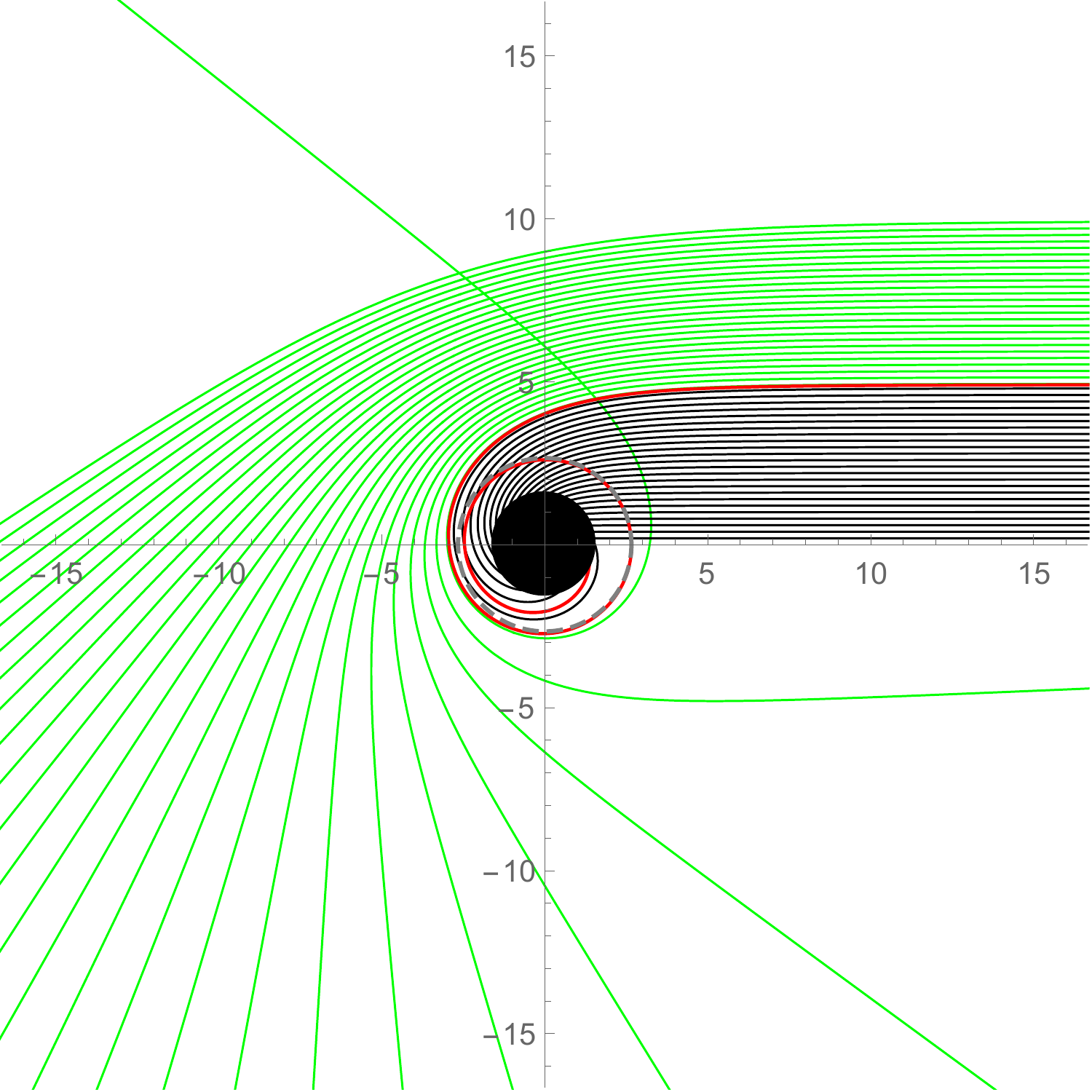}}
\caption{The trajectories of light ray for different values of $\omega$ in the polar coordinates $(r, \phi)$. }\label{figure2}
\end{figure}

Next, let's turn attentions on the trajectories of the light ray.
Using Eqs.(\ref{eq101}) and (\ref{eq102}), the concrete expression of motion equation of photon can be expressed as
\begin{align}\label{eq15}
\frac{{dr}}{d \phi }=\pm r^2\sqrt{\frac{1}{b_{ c}^2} - \frac{1}{r^2}\left[ 1 + \frac{1}{2\omega^2} \left(r^2 - \sqrt{r^4 + 8 M r \omega^2 }  \right) \right]}.
\end{align}
By introducing a new parameter $u=1/r$, then it yields
\begin{align}\label{eq16}
\mathcal{R}(u,b)\equiv \frac{d u}{d \phi }=\sqrt{\frac{1}{b_{ c}^2} - u^2\left[ 1 + \frac{1}{2\omega^2} \left(\frac{1}{u^2} - \sqrt{\frac{1}{u^4} +  \frac{8 M  \omega^2}{u} }  \right) \right]}
\end{align}
With the help of the ray-tracing code, the trajectory of light ray for different values of $\omega$ are shown in Figure \ref{figure2}.

In Figure \ref{figure2}, the black hole is shown as a black disk and the dashed grey circle is the photon orbit. The spacing of impact parameter $b_c$ has been fixed to 0.2.
Obviously, one can see that the black lines all dropped into the black hole, which means those lines correspond to $b_c < b_p$ and the Region 3 in Figure \ref{figure1}. For the red line, it revolve around the black hole, thereby this line corresponds to $b_c = b_p$ and the Region 2 in Figure \ref{figure1}. Also, the green lines all deflected around the black hole, and then moved to infinity. Those lines correspond to $b_c >b_p$ and the Region 1 in Figure \ref{figure1}. More importantly, we find that the black disk of the case $\omega = 0.1$ is bigger than that of the case $\omega = 0.8$, which implies the size of black hole increases with the decrease of parameter $\omega$.

\section{ Shadows and rings of KS black hole }\label{sec3}
Considering the fact that there are always various matters around the black hole in the real universe, so it is necessary to further study the observational appearance in this situation. For simplicity, we in this section take an optically and geometrically thin disk accretion as an example to study the shadows and rings of KS black hole.

\subsection{Direct emission, Lensed ring and photon ring}\label{sec31}
In 2019, Gralla, Holz and Wald have reported that there are various cases of emission from the region near the black hole for a distant observer\cite{Wald}. According to the definition of the total number of orbits $ n = \phi /{2 \pi}$ , the trajectories of light rays emitted from the north pole direction can be divided into three case\footnote{Where, $n$ is a function of the impact parameter $b_c$.}. \emph{Case A}: namely direct emission, when the number of orbits $n<3/4$, the trajectories of light rays will intersect with the equatorial plane only once. \emph{Case B}: namely Lensing ring, when the number of orbits $3/4<n<5/4$, the trajectories of light rays will intersect with the equatorial plane at least twice. \emph{Case C}: namely photon ring, when the number of orbits $n>5/4$, the trajectories of light rays will intersect with the equatorial plane at least three times. In the context of KS black hole, by taking the $\omega = 0.1$ and $\omega = 0.8 $ as two example, the regions of the direct emission, lensing ring and photon ring with respect to the impact parameter $b_c$ have been shown in Eqs. (\ref{Eq119}) and (\ref{Eq120}), where the value of $M$ has been set to 1.
\begin{align} \label{Eq119}
 \omega=0.1 \Rightarrow
    \begin{cases}
   \text{Direct emission}: n<3/4,~~~
    b_c < 5.00984  ~~ \text{and} ~~  b_c> 6.16591 \\
     \text{Lensing ring}: 3/4<n<5/4,~~~
   5.00984<b_c< 5.18382 ~~  \text{and}   ~~ 5.22435<b_c<6.16591  \\
       \text{Photon ring}: n>5/4,~~~
     5.18382< b_c < 5.22435
    \end{cases}
\end{align}
\begin{align}  \label{Eq120}
  \omega=0.8 \Rightarrow
    \begin{cases}
   \text{Direct emission}: n<3/4,~~~
    b_c < 4.56591  ~~ \text{and} ~~  b_c >6.05939 \\
     \text{Lensing ring}: 3/4<n<5/4,~~~
  4.56591<b_c<4.88277 ~~  \text{and}   ~~ 4.97204<b_c< 6.05939  \\
       \text{Photon ring}: n>5/4,~~~
     4.88277< b_c < 4.97204
    \end{cases}
\end{align}

\begin{figure}[!h]
\centering
\includegraphics[width=.55\textwidth]{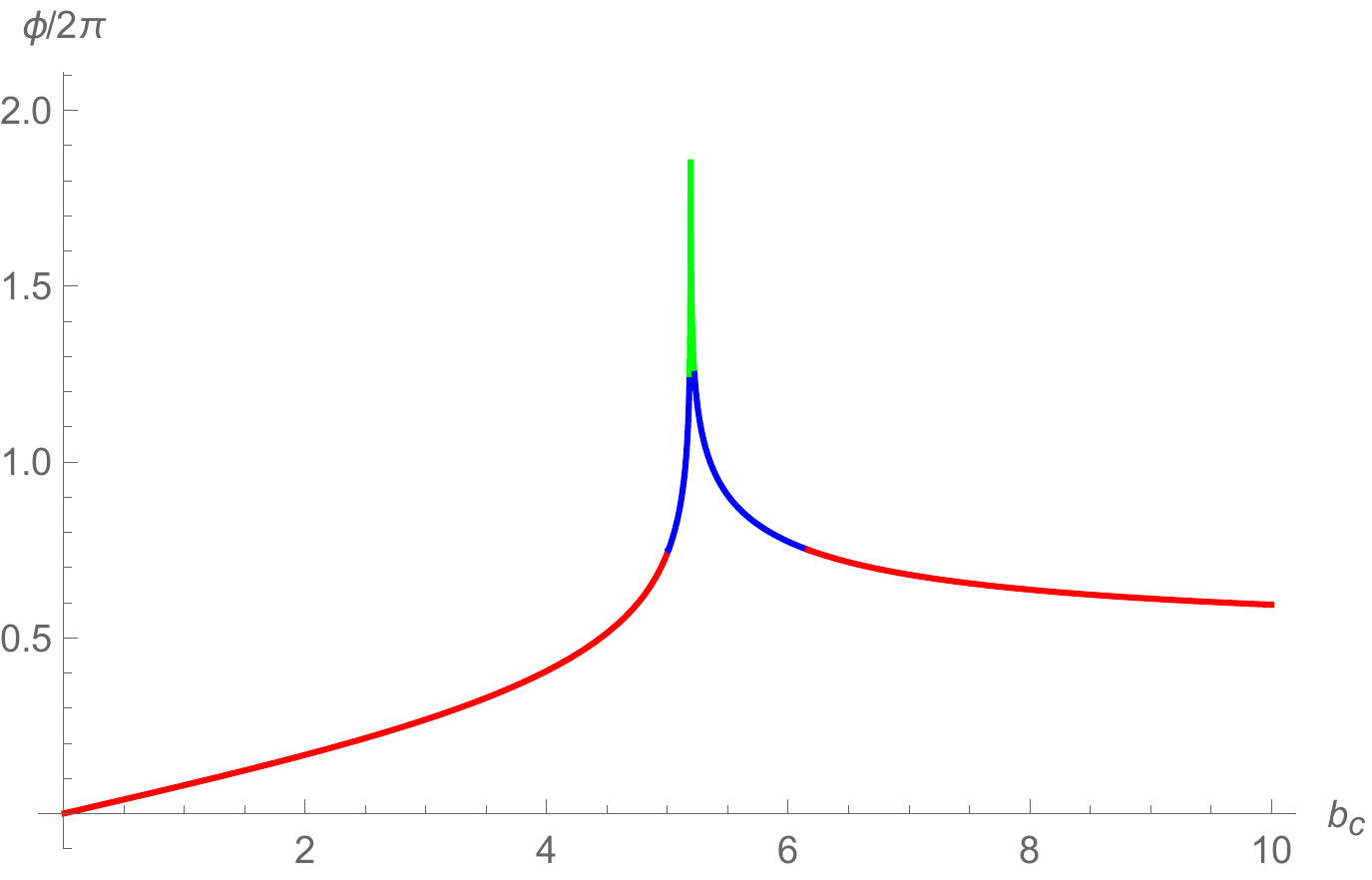}
\hfill
\includegraphics[width=.43\textwidth]{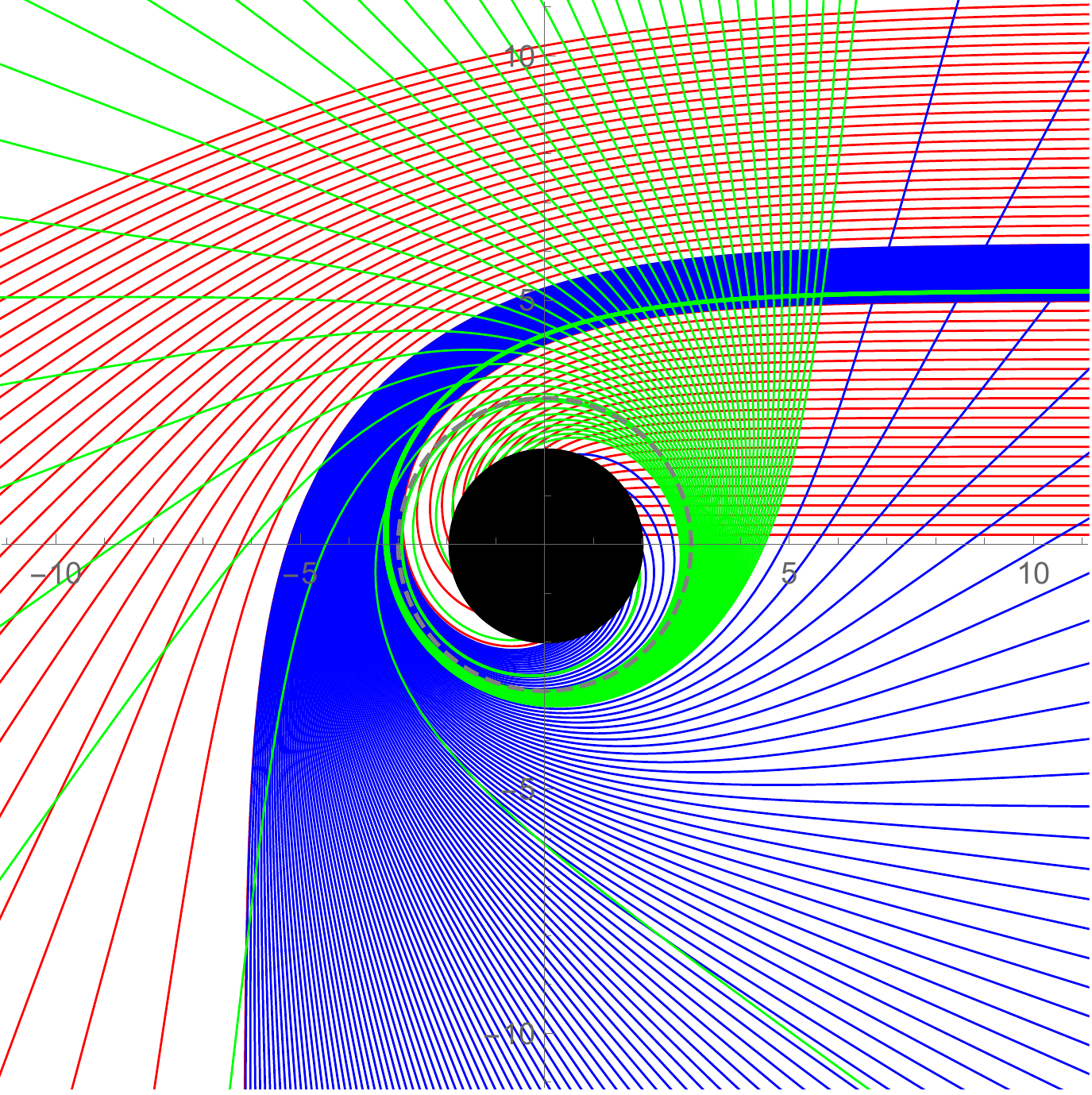}
\includegraphics[width=.56\textwidth]{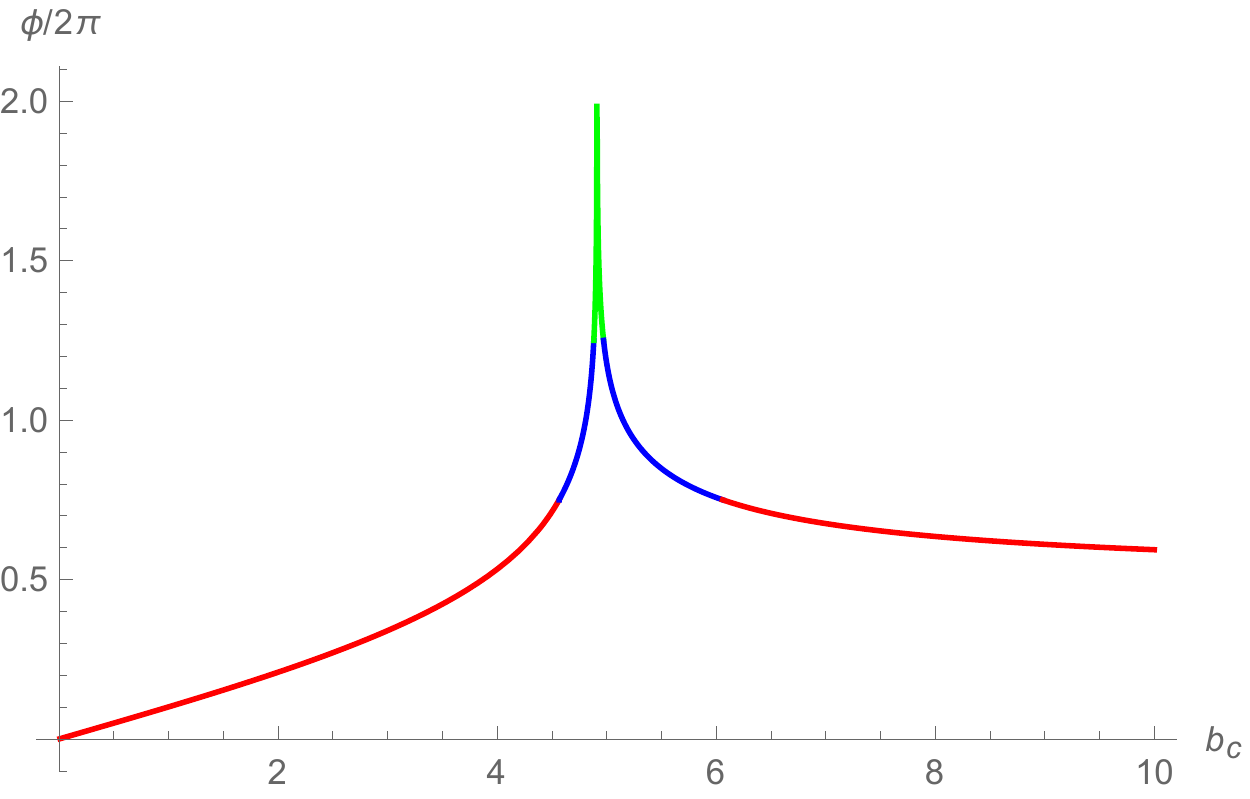}
\hfill
\includegraphics[width=.43\textwidth]{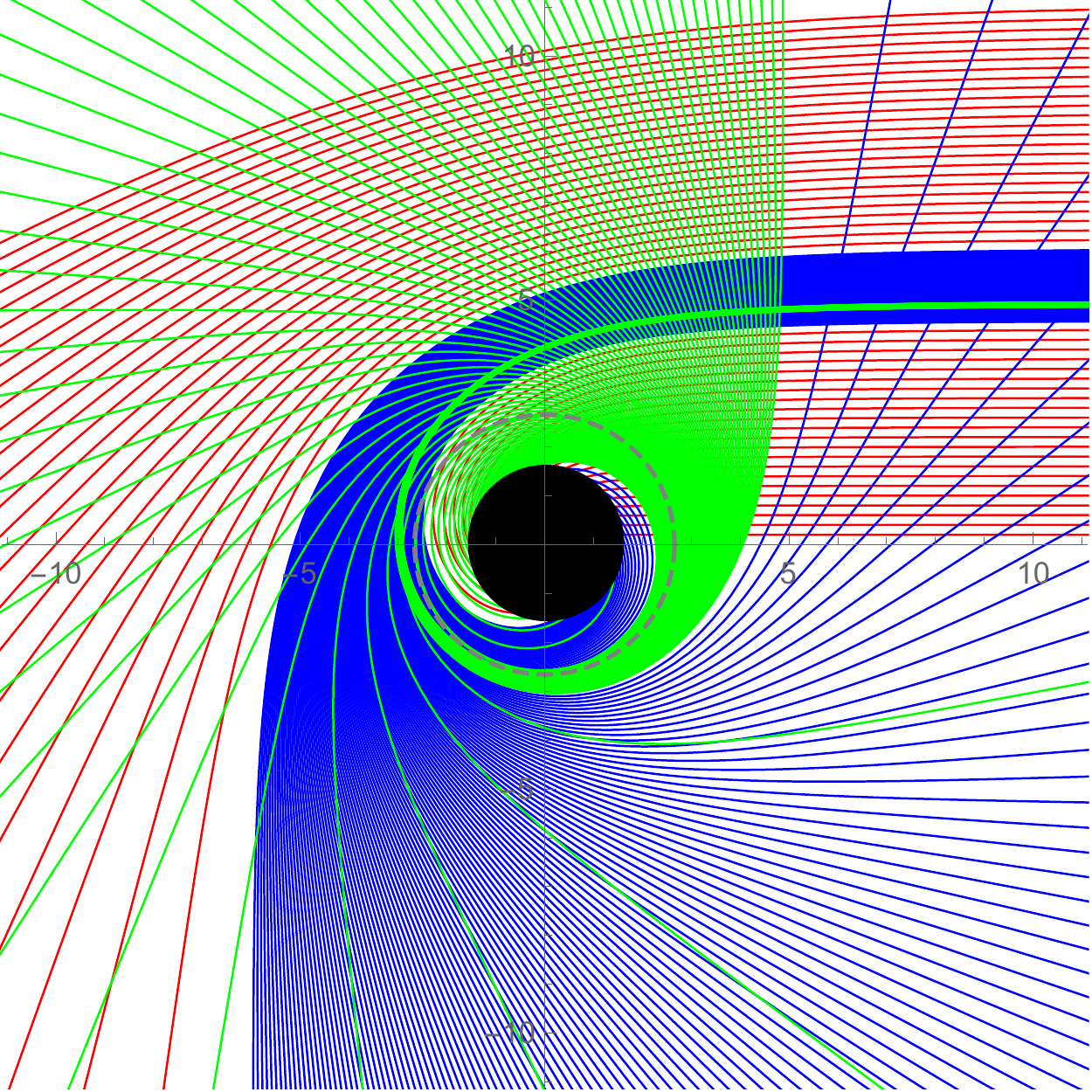}
\caption{\label{fig3} Behavior of photons in the KS black hole as a function of the impact parameter $b_c$.}
\end{figure}

From Eqs. (\ref{Eq119}) and (\ref{Eq120}), it is easy to see that the regions of the direct emission, lensed ring and photon ring are all smaller and smaller with the increase of parameter $\omega$. In addition, to present the direct emission, lensed ring and photon ring more intuitively, we have plotted Figure \ref{fig3} for different values of $\omega$.

For the left column, it represents the fractional number of orbits $n$ versus the impact parameter $b_c$. And in the polar coordinates ($b_c, \phi$), the corresponding photon trajectories of the KS black hole are shown in the right column. For the first line, it corresponds to the case of $\omega=0.1$, and the case $\omega=0.8$ are presented in the second line. The spacings of $b_c$ are 1/5, 1/100, 1/1000 for the direct (red line), lensed (blue line) and photon (green line) ring trajectories, respectively. The dashed grey line is the photon orbit and the black solid disk represents the KS black hole. From Figure \ref{fig3}, it is obvious that the parameter $\omega$ has an influence that cannot be ignored on the trajectories of light rays, which might be acted as an effective feature to distinguish the KS black hole from the Schwarzschild black hole.

\subsection{Observed specific intensities and transfer functions}
\begin{figure}[!h]
\centering
\subfigure[$\omega=0.1$]{
\includegraphics[scale=0.45]{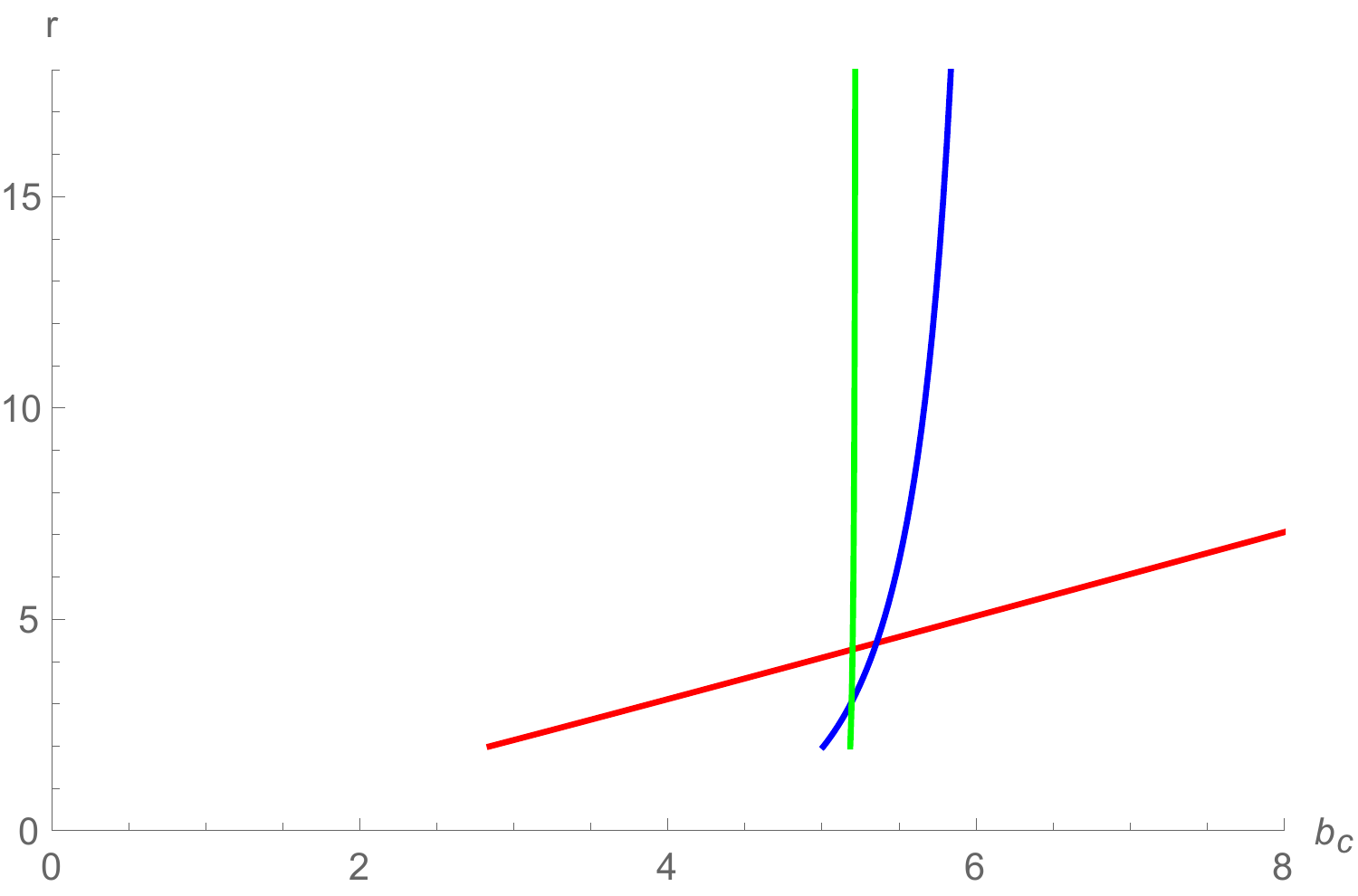}}
\subfigure[$\omega=0.8$]{
\includegraphics[scale=0.487]{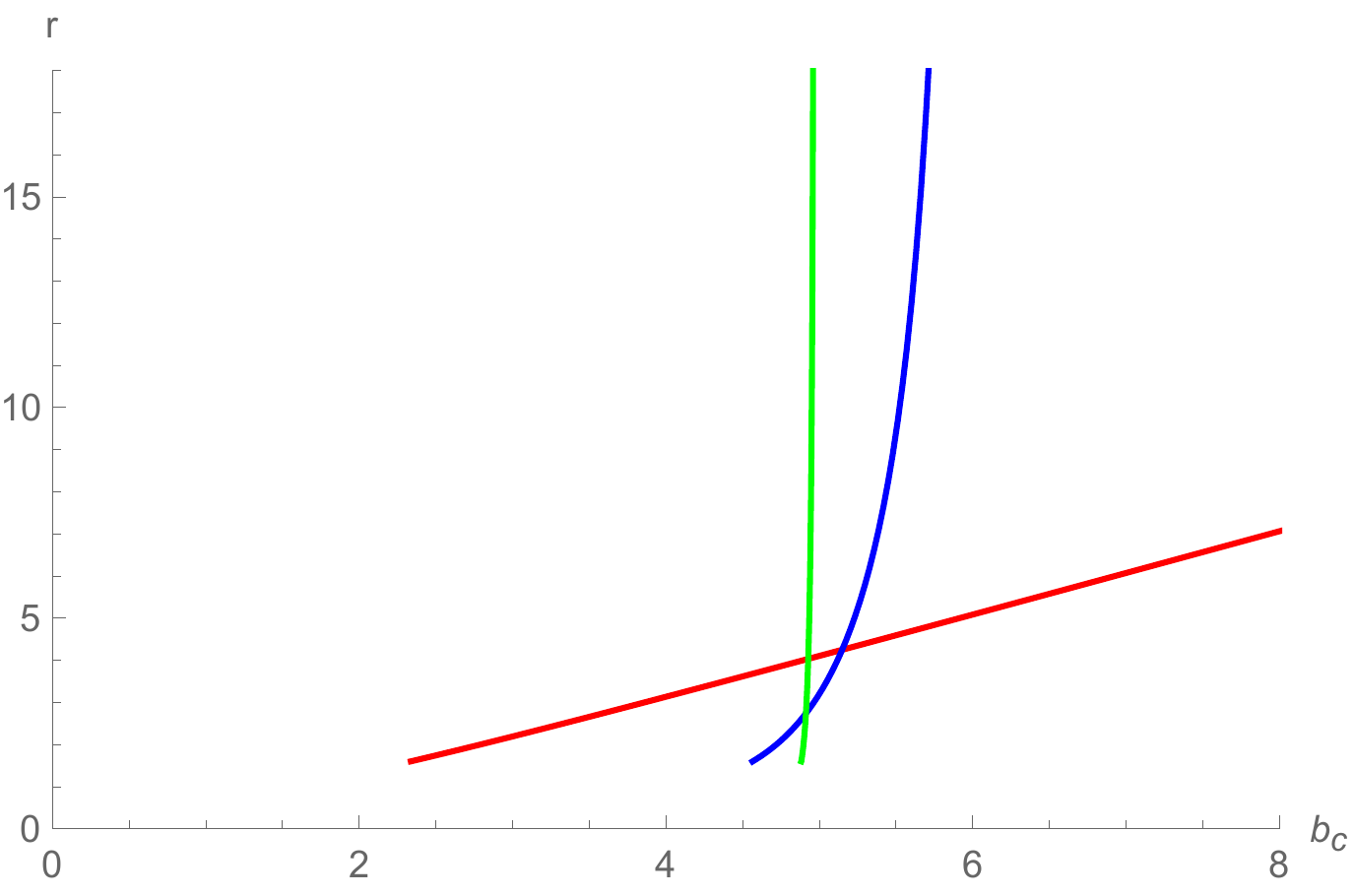}}
\caption{  The first three transfer functions for KS black hole. }\label{fig4}
\end{figure}

In the next subsection, when an optically and geometrically thin-disk accretion surrounded the black hole, we will continue to investigate the observed specific intensity of the thin disk accretion. For simplicity, we assume that the thin disk is located at the rest frame of static worldlines, and the photons emitted from it is isotropic. Meanwhile, we further consider a situation that the thin disk lies in the equatorial plane of the KS black hole and the static observer locates at the north pole. Under those considerations, the observed specific intensity can be expressed as
\begin{align}\label{eq17}
I_{obs}(r)= \left[1 + \frac{1}{2\omega^2} \left(r^2 - \sqrt{r^4 + 8 M r \omega^2 }  \right)\right]^{3/2} I_{emi}(r),
\end{align}
where $I_{obs}(r)$, $I_{emi}(r)$ are the observed specific intensity with frequency $\nu$ and the emitted specific intensity with frequency $\nu_e$. So, by integrating specific intensity with frequencies $\nu$, the total specific intensity can be obtained, which is
\begin{align}
I(r) &= \int I_{obs}(r)d\nu =\int \left[1 + \frac{1}{2\omega^2} \left(r^2 - \sqrt{r^4 + 8 M r \omega^2 }  \right)\right]^2 I_{emi}(r) d\nu_e \label{eq181}\\
&= \left[1 + \frac{1}{2\omega^2} \left(r^2 - \sqrt{r^4 + 8 M r \omega^2 }  \right)\right]^2 I_{em}(r).\label{eq182}
\end{align}
Here, $I_{em}(r) = \int I_{emi}(r) d\nu_e$ denotes the total emitted specific intensity, and the relationship $\nu = f(r)^{1/2} \nu_e$ has been used in above equation.
As described in subsection \ref{sec31}, we know that the light rays near the black hole will intersect with the equatorial plane, hence that means it will intersect with the thin disk lied in the equatorial plane of the KS black hole. For each intersection, the light ray will pick up brightness from the disk emission. Therefore, considering the light rays emitted from thin disk, when $3/4< n <5/4$ (Lensed rings, which is the blue line in Figure \ref{fig3}), the light ray will produce a intersection with the disk at the opposite side of it. Then, an additional brightness from this intersection will be got by the light ray. When $n > 5/4$ (photon rings, which is the green line in Figure \ref{fig3}), the light ray will produce a intersection with the disk at the front side of it once again. This naturally give rise to an additional brightness of the light ray once again. Combined with those facts, the observed intensity should be a sum of those intensities from each intersection. So, we have
\begin{align}\label{eq19}
I(r) = \sum_n f(r)^2 I_{em}(r)|_{r=r_n(b_c)},
\end{align}
where $r_n(b_c)$ is defined as the transfer function, which represents the radial location of the $n$th intersection with the disk.
It should be noted that we do not consider the absorption of light rays' brightness since it will abate the observation intensity.
The transfer function $r_n(b_c)$ does not only describe the location $r$ that the light ray with impact parameter $b_c$ hit the disk on, but also imply the demagnification factor\footnote{This factor is described by the slope $dr/{db_c}$}.
For different choice of parameter $\omega$, we have plotted Figure \ref{fig4} to clearly show the relationship between the transfer functions $r_n(b_c)$  and the impact parameter $b_c$.

In Figure \ref{fig4}, the red line, which corresponds to the first($n=1$) transfer function, represents the direct image of the disk. In the case, the direct image profile is the redshifted source profile since its slope is approximatively equal to 1. The blue line, which corresponds to the second($n=2$) transfer function, represents the lensing ring (with inclusion of the photon ring). In this situation, the image of the back side of the disk will be demagnified, because its slope is much greater than 1. The green line, which corresponds to the third($n=3$) transfer function, represents the photon ring. In this sense, due to the infinity of the slope, the image of the front side of the disk will be extremely demagnified. On the other hand, one can see that the parameter $\omega$ also has an important influence that cannot be ignored on the first three transfer functions.

\subsection{Observational features of Direct emissions and rings}
\begin{figure}[h]
\centering 
\includegraphics[width=.35\textwidth]{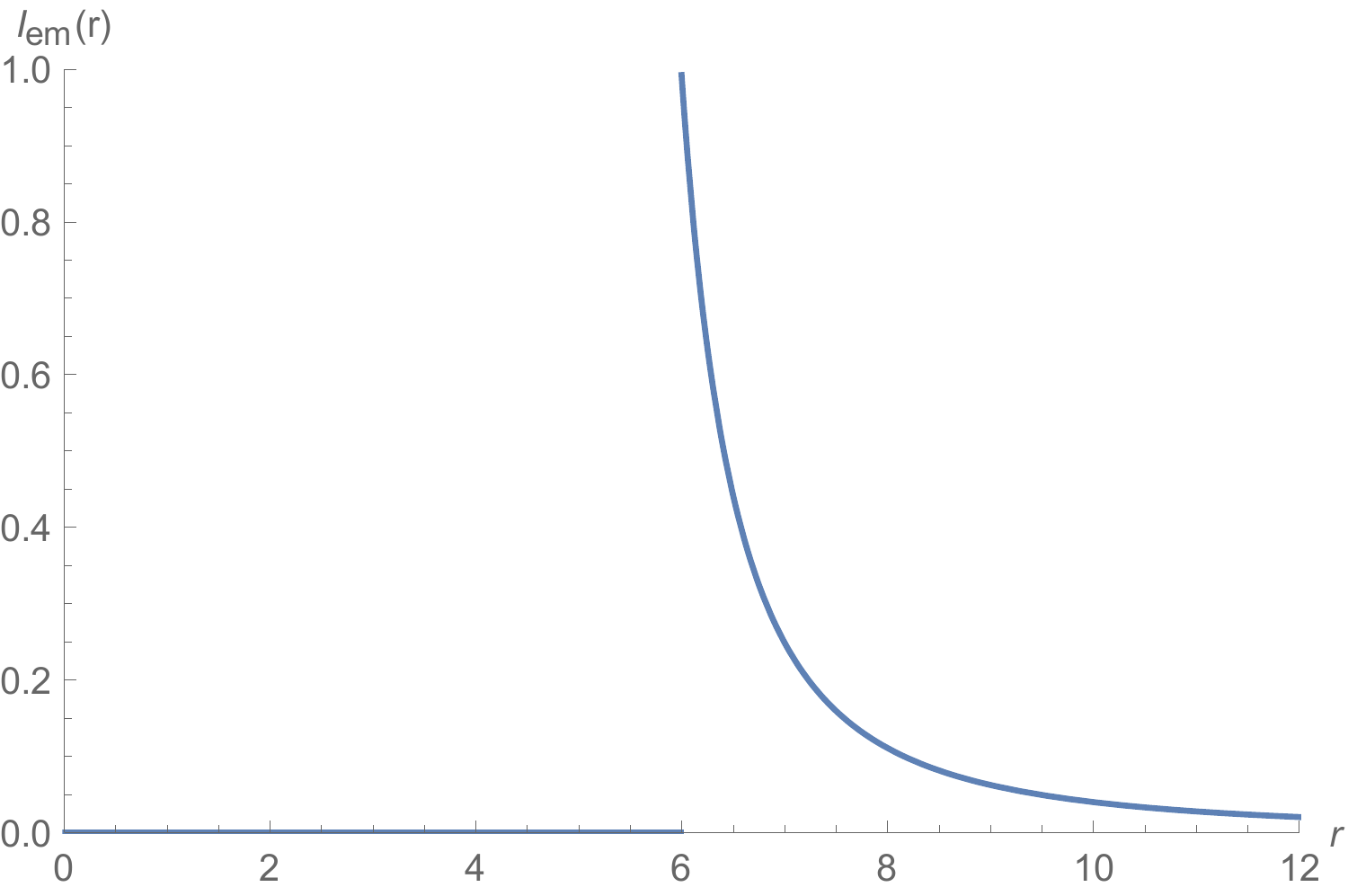}
\includegraphics[width=.35\textwidth]{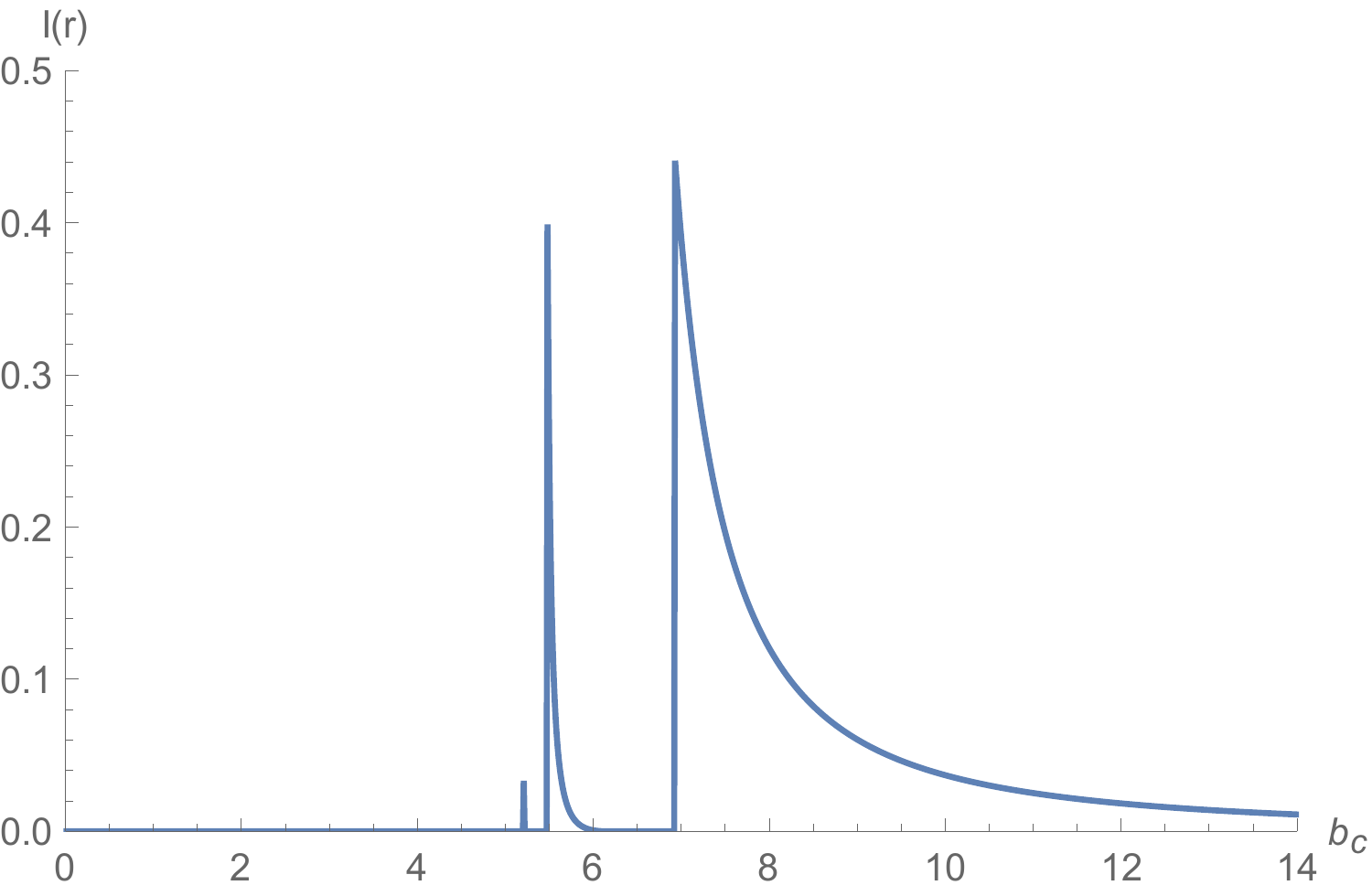}
\includegraphics[width=.28\textwidth]{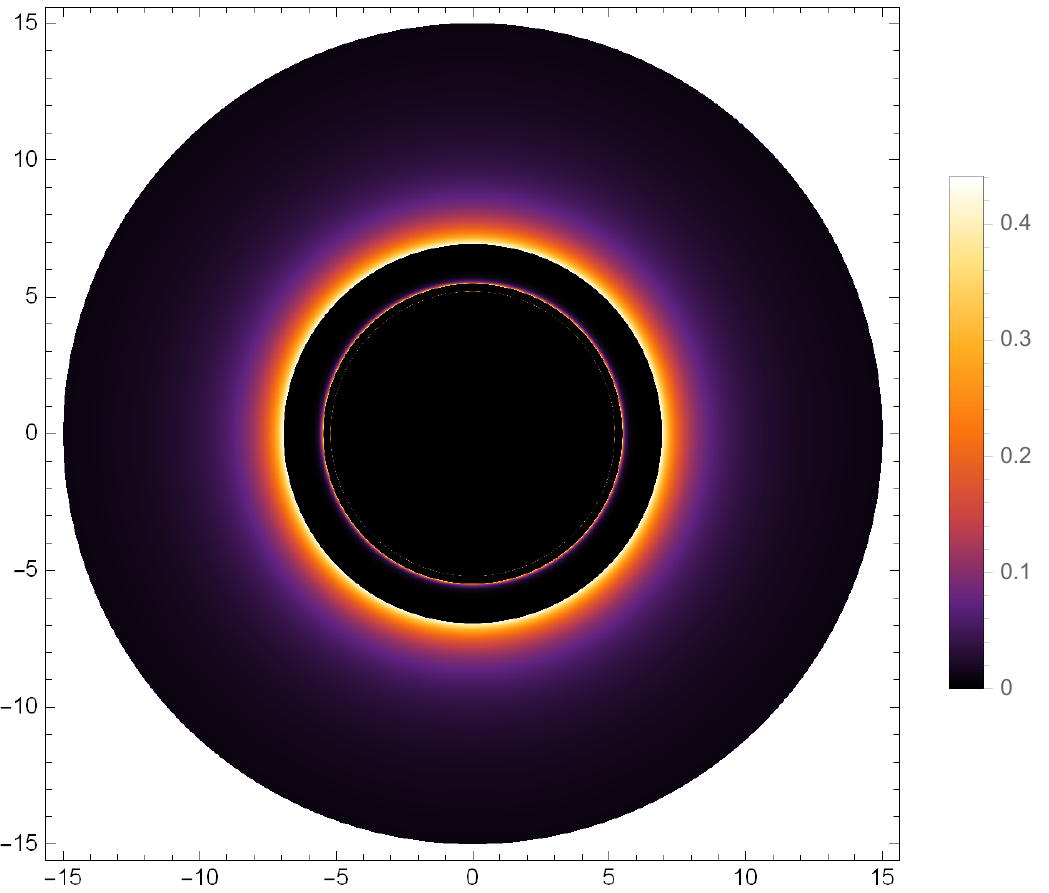}
\includegraphics[width=.35\textwidth]{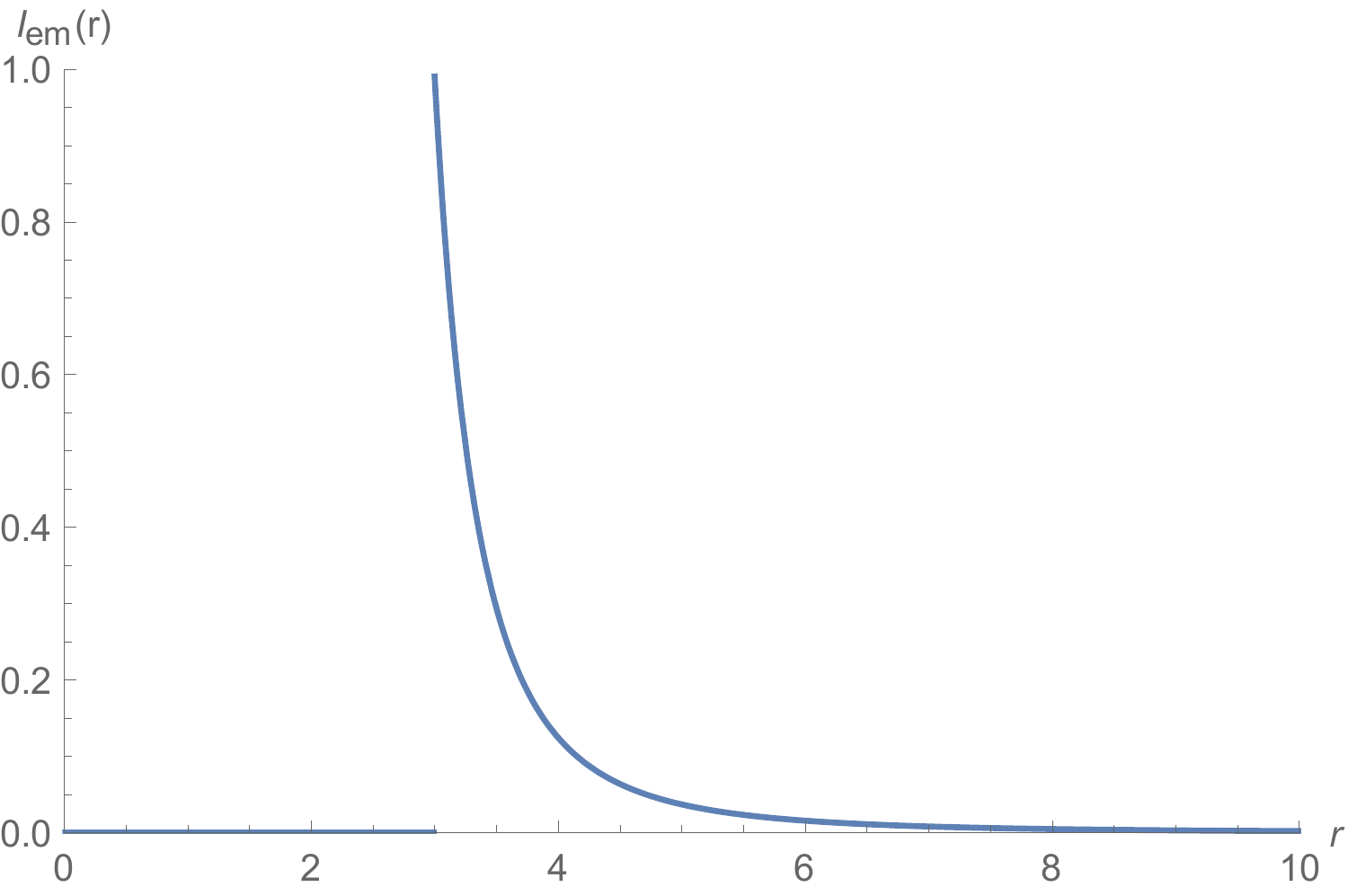}
\includegraphics[width=.35\textwidth]{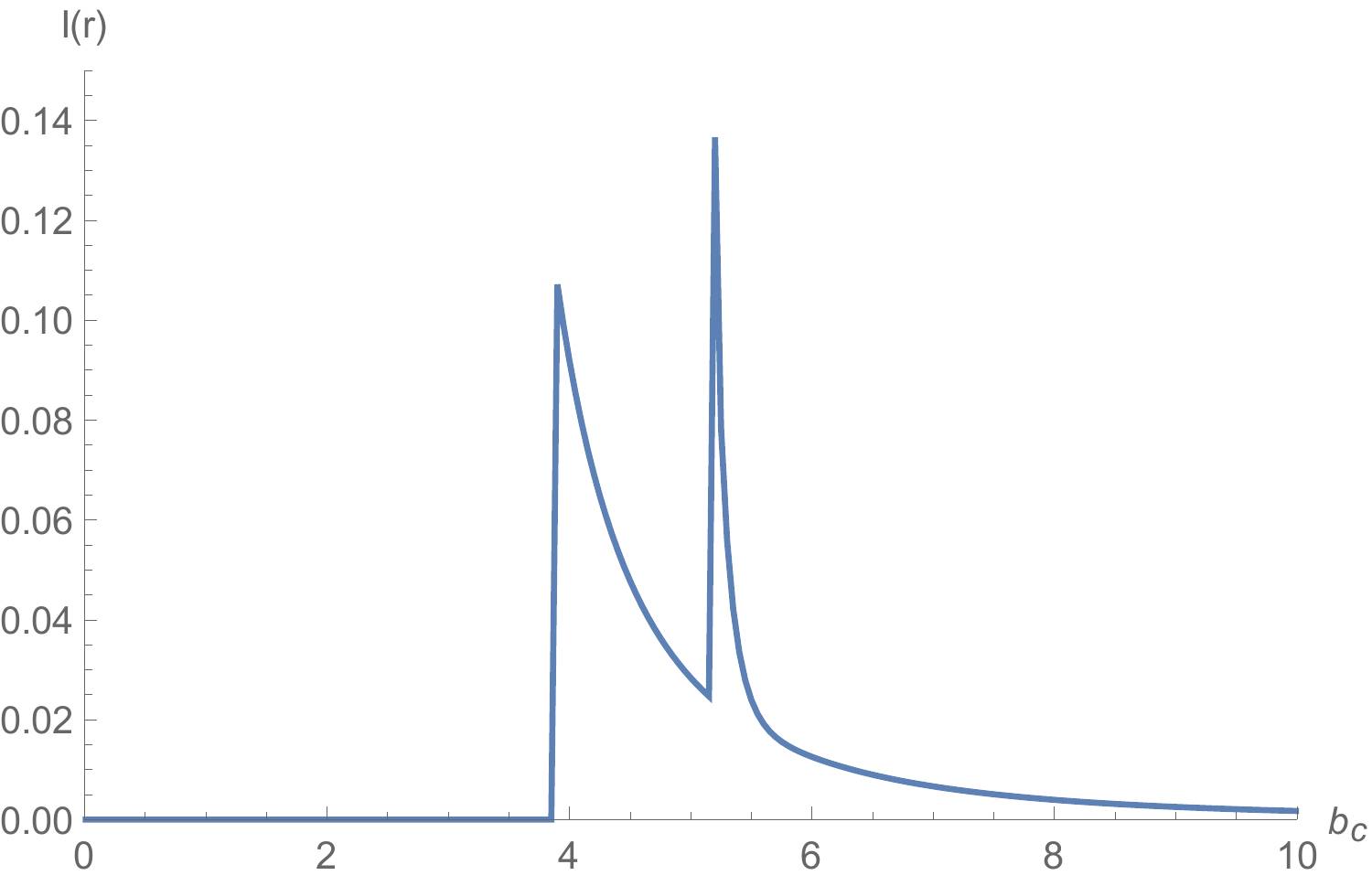}
\includegraphics[width=.28\textwidth]{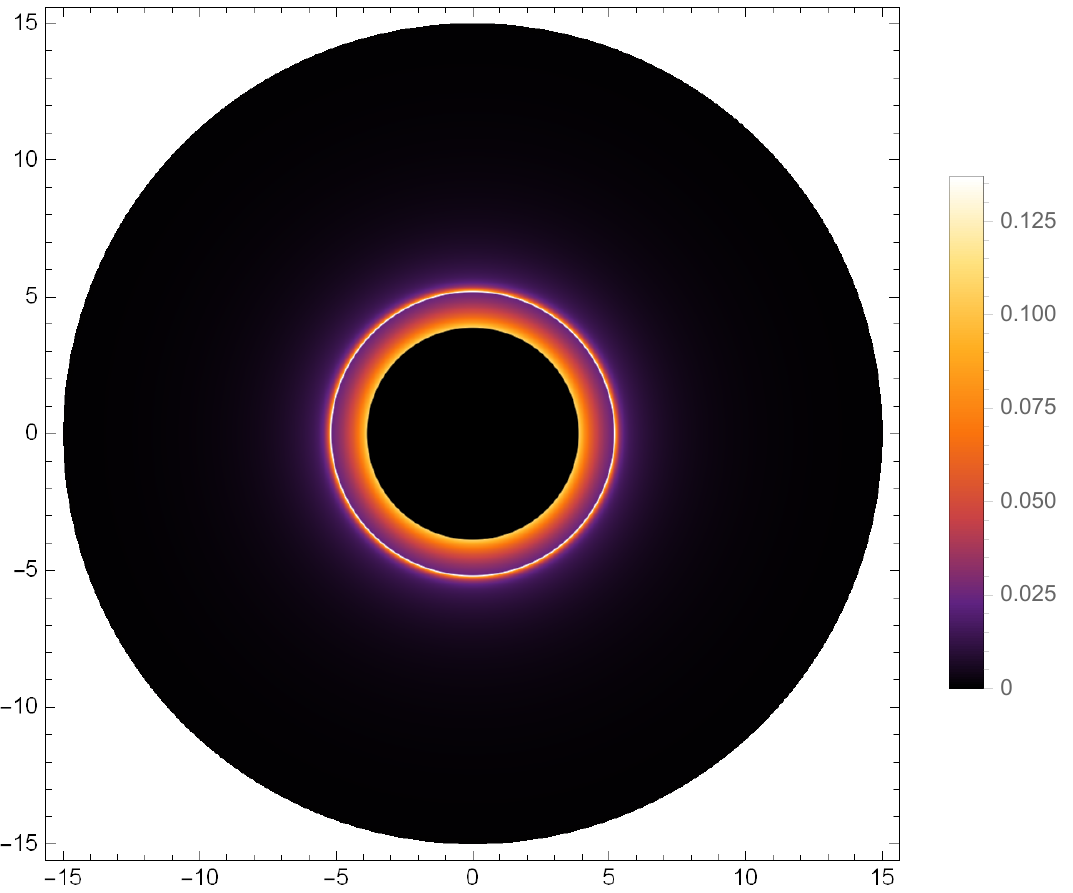}
\includegraphics[width=.35\textwidth]{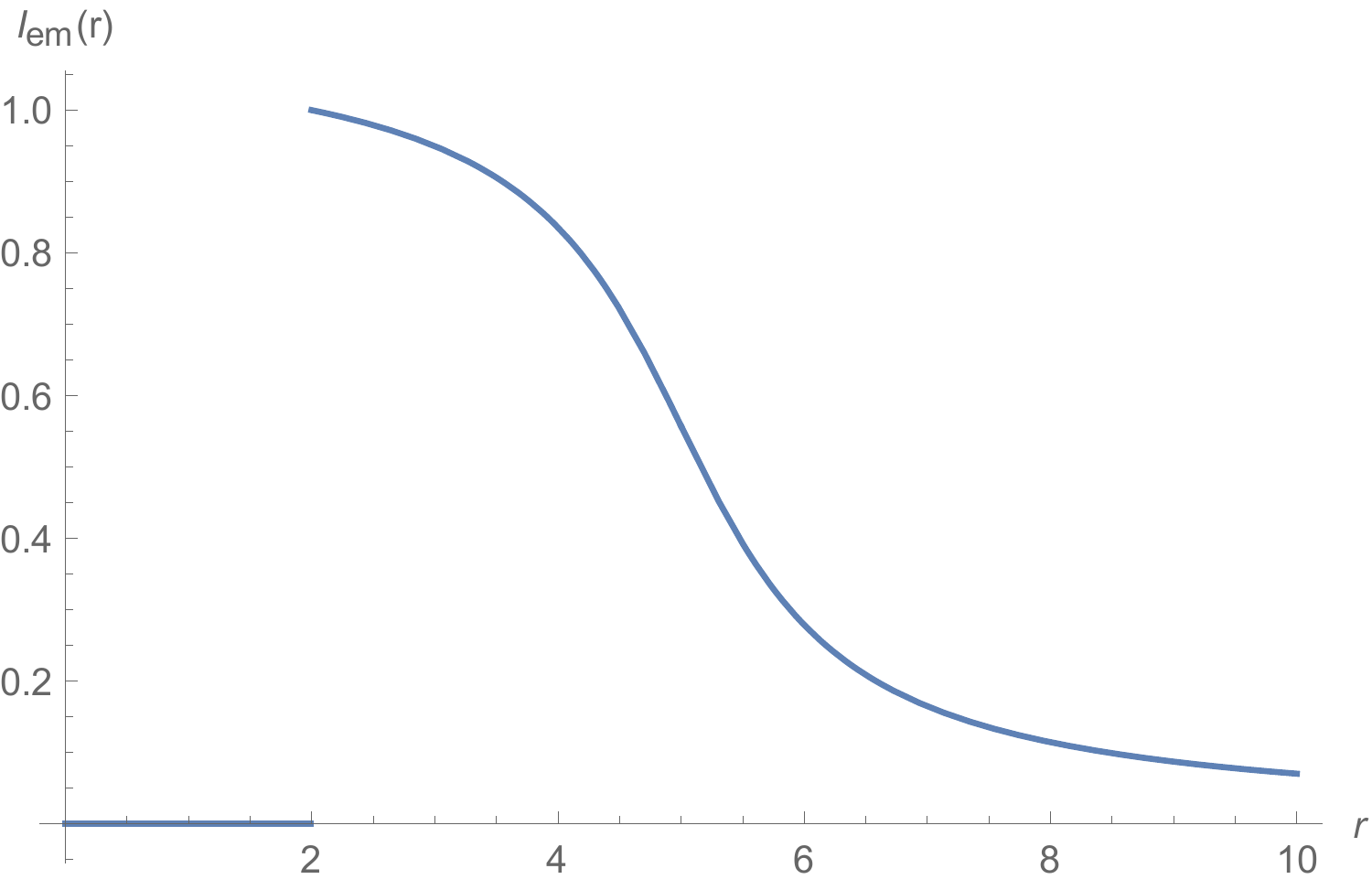}
\includegraphics[width=.3525\textwidth]{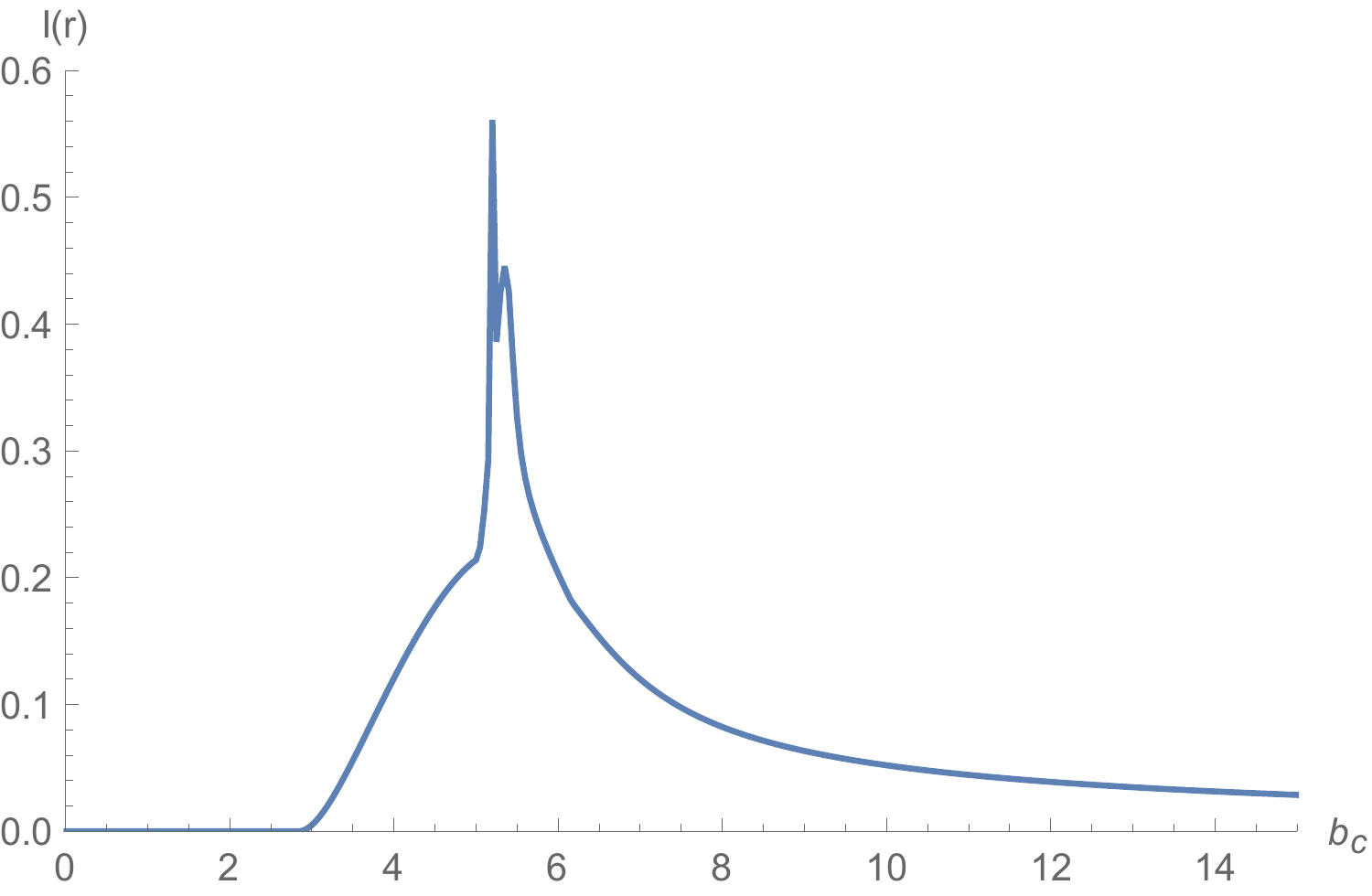}
\includegraphics[width=.2775\textwidth]{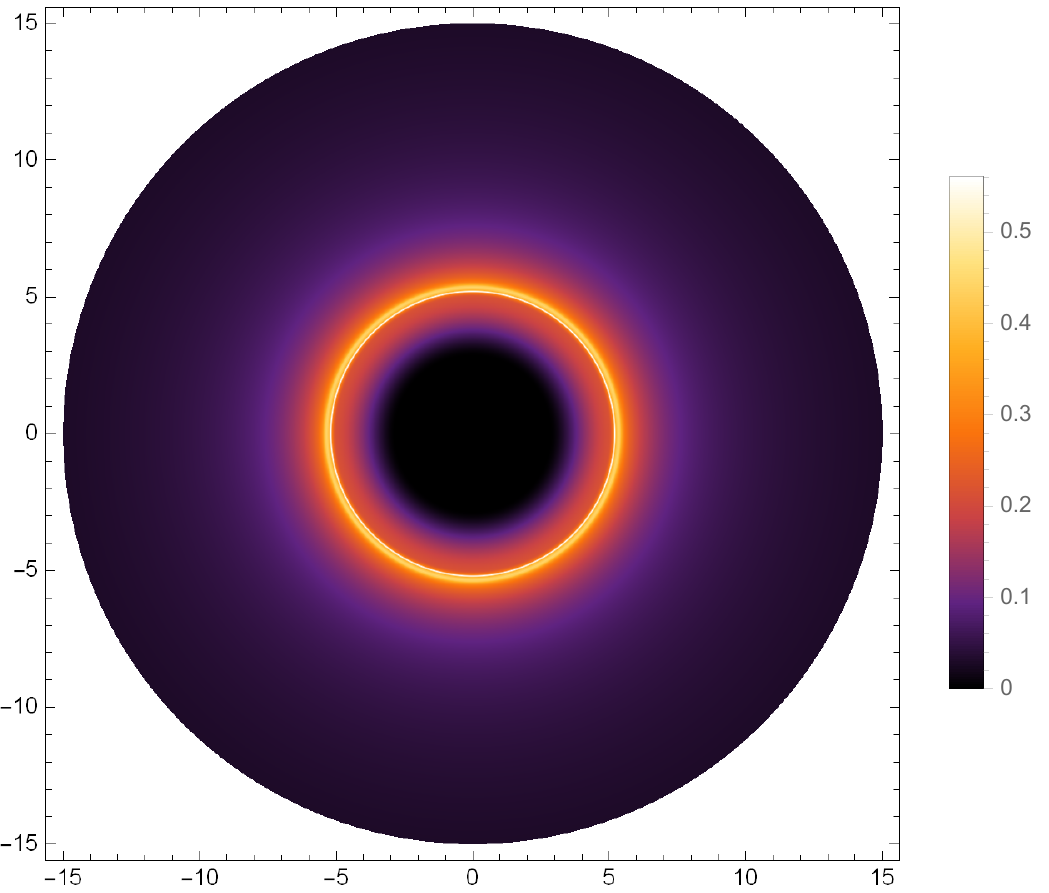}
\caption{\label{fig5}  Observational appearances of a geometrically and optically thin disk with different profiles near black hole. }
\end{figure}

\begin{figure}[h]
\centering 
\includegraphics[width=.35\textwidth]{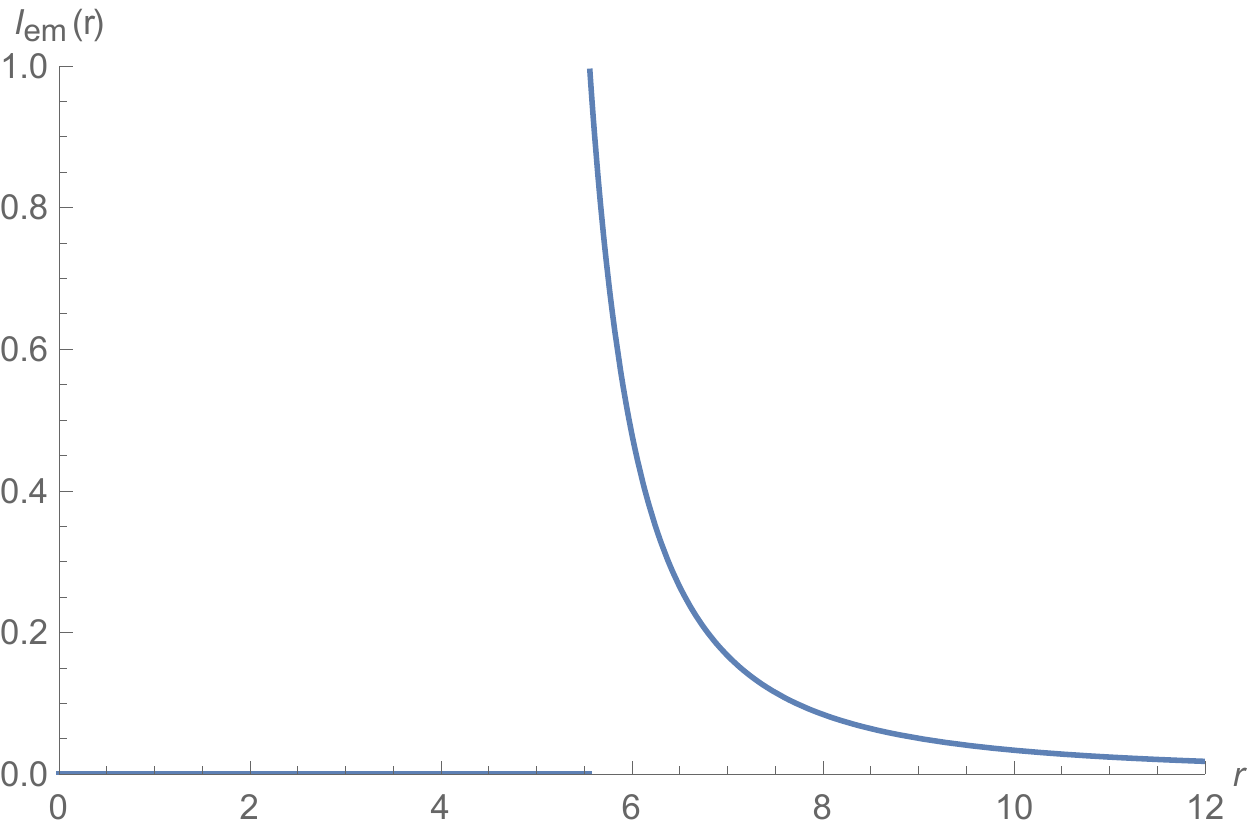}
\includegraphics[width=.35\textwidth]{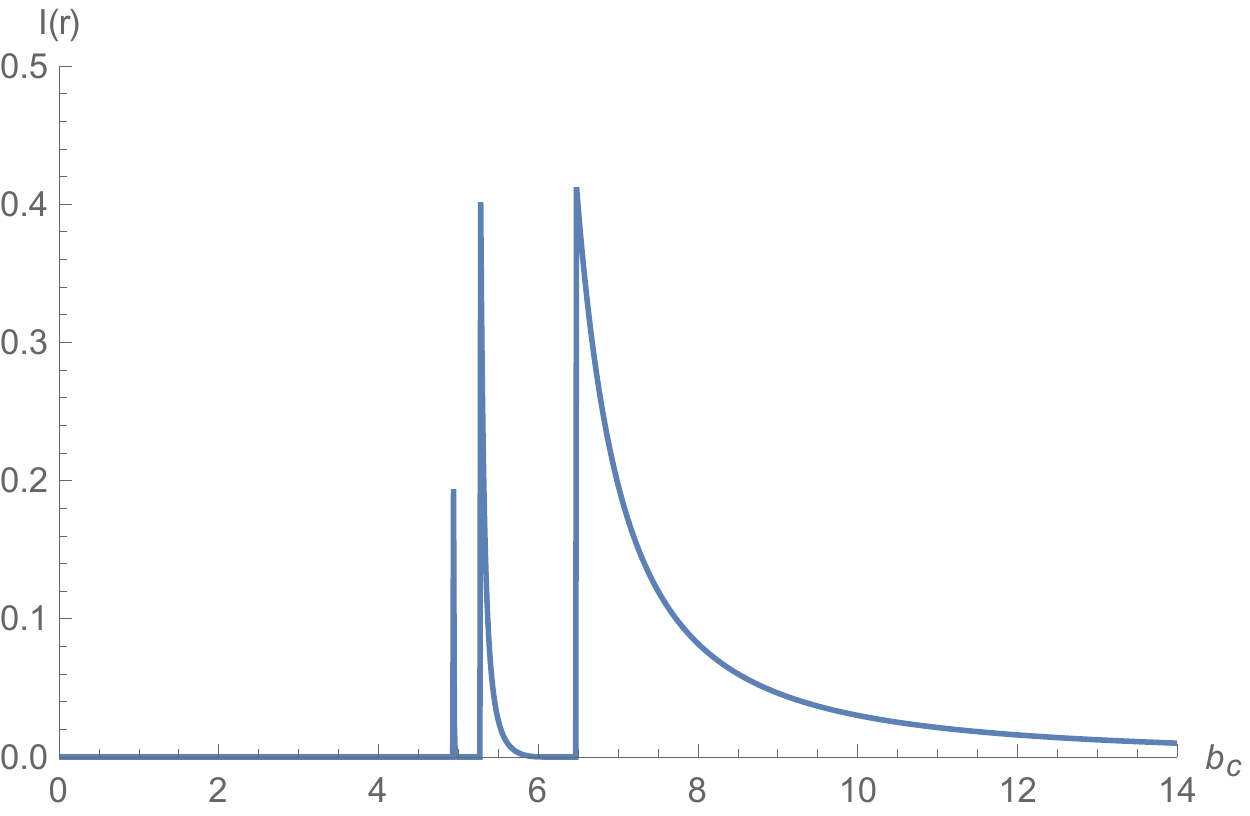}
\includegraphics[width=.28\textwidth]{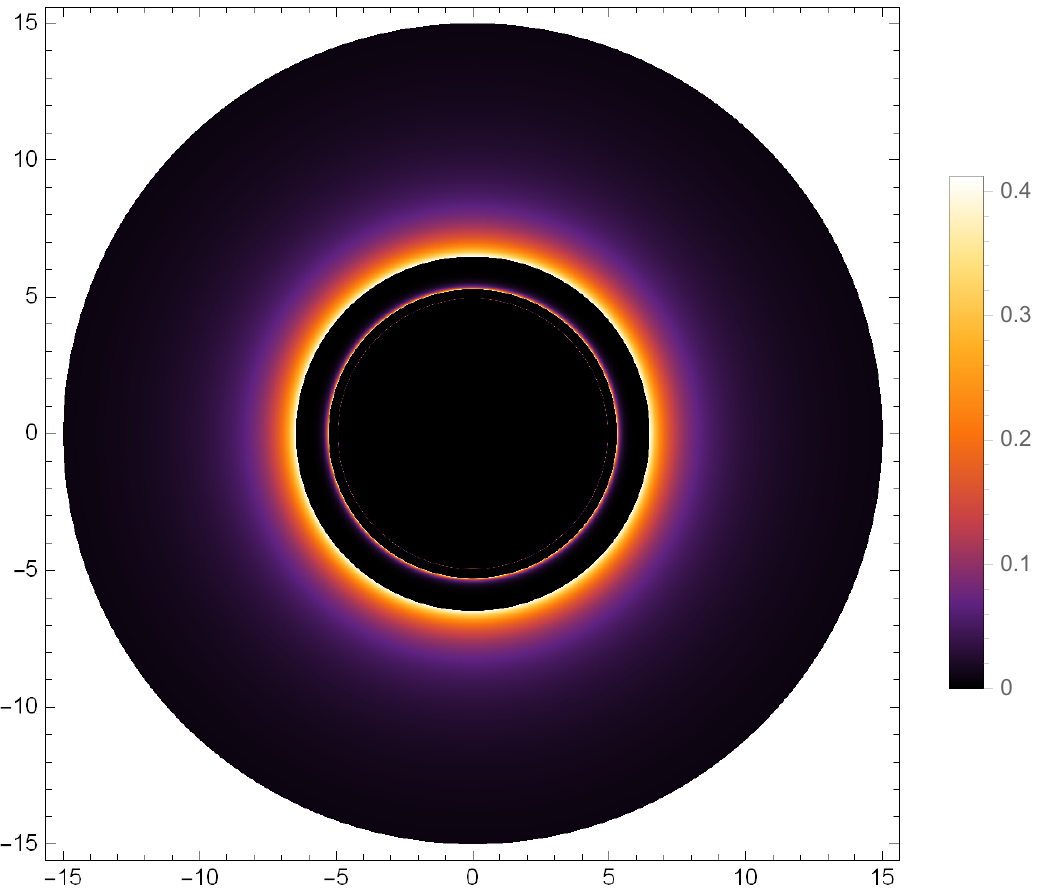}
\includegraphics[width=.35\textwidth]{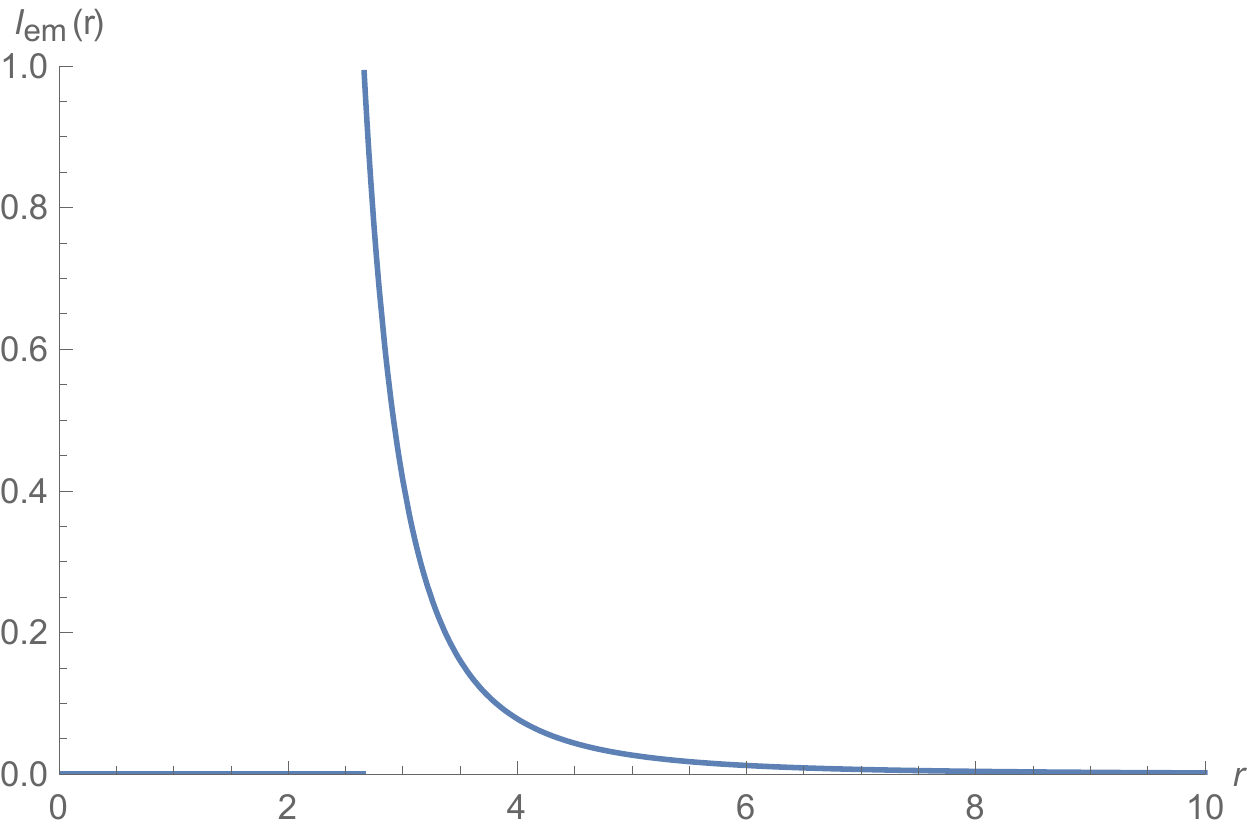}
\includegraphics[width=.35\textwidth]{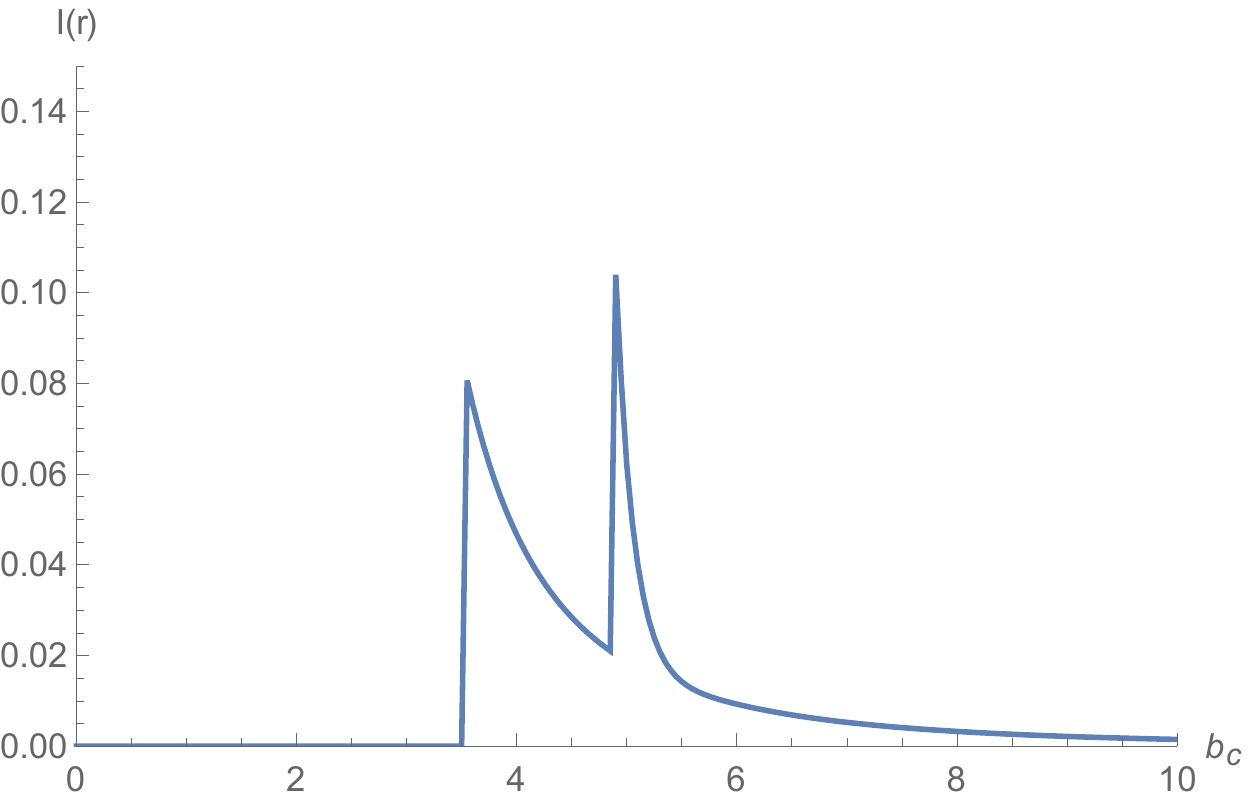}
\includegraphics[width=.28\textwidth]{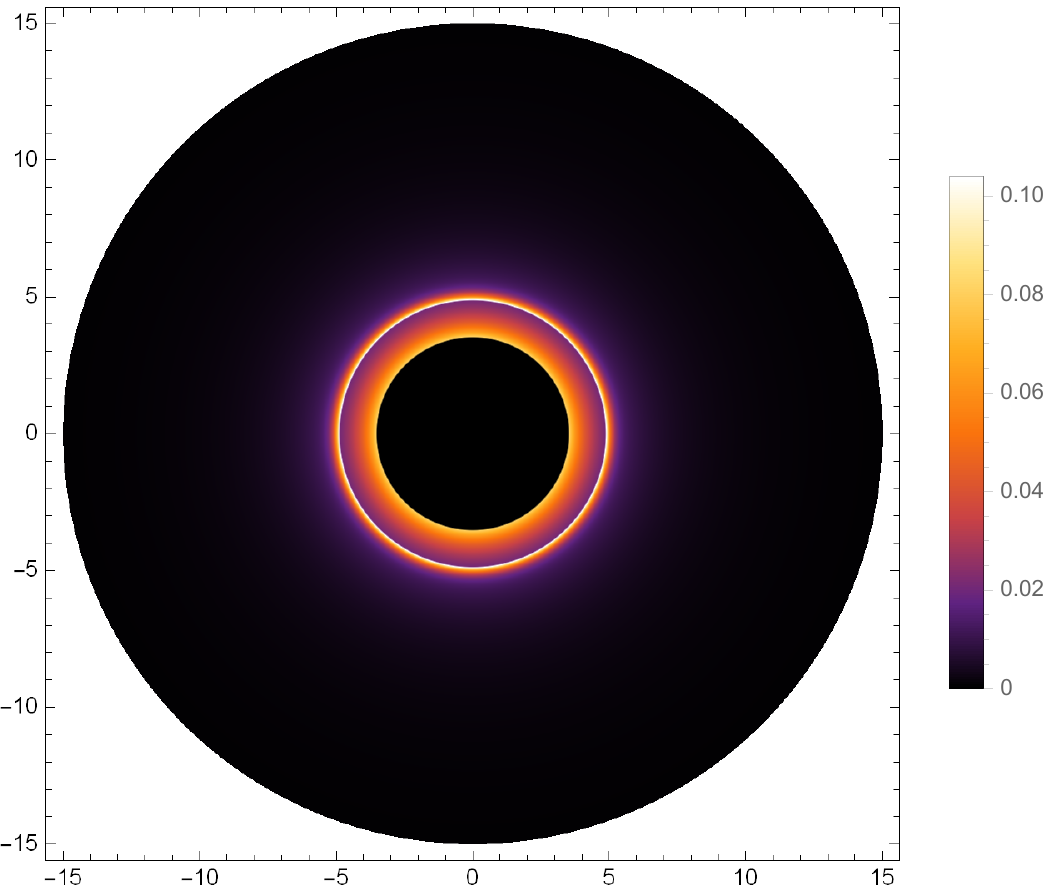}
\includegraphics[width=.35\textwidth]{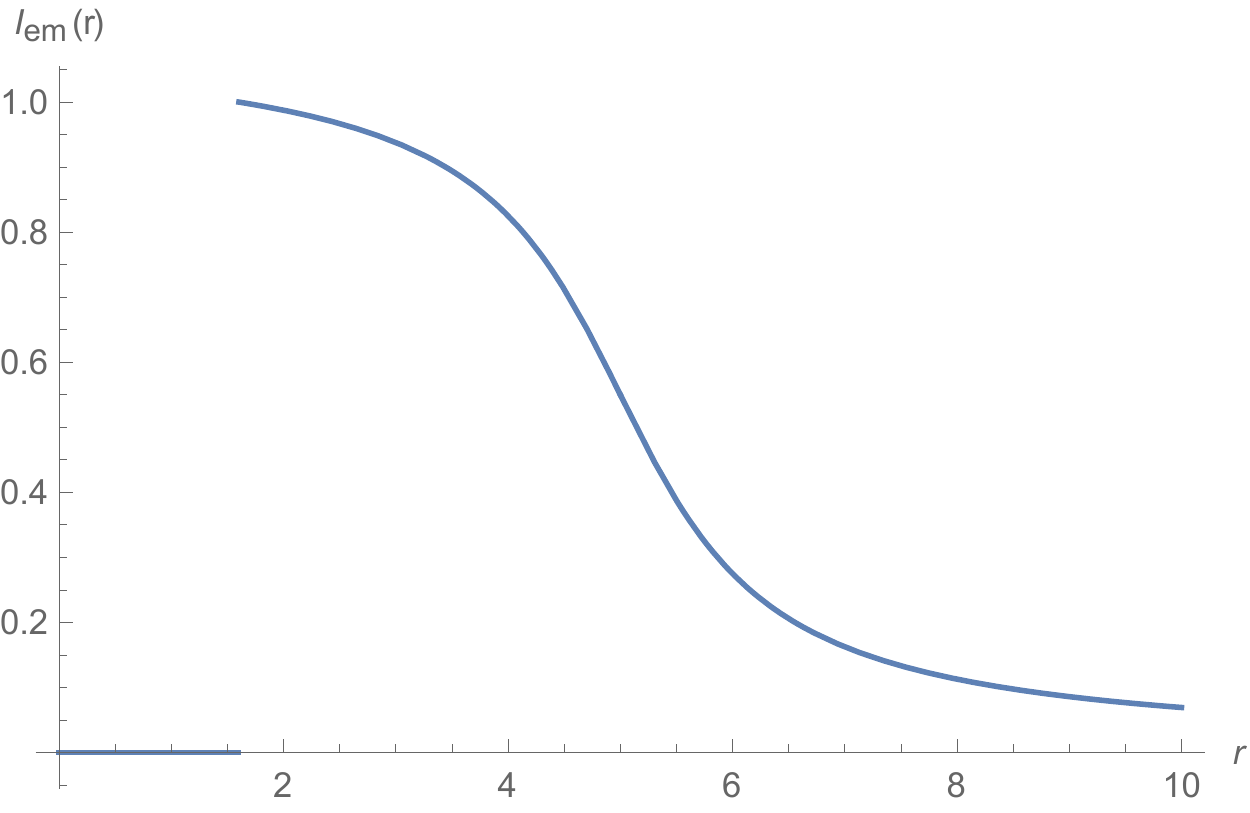}
\includegraphics[width=.3525\textwidth]{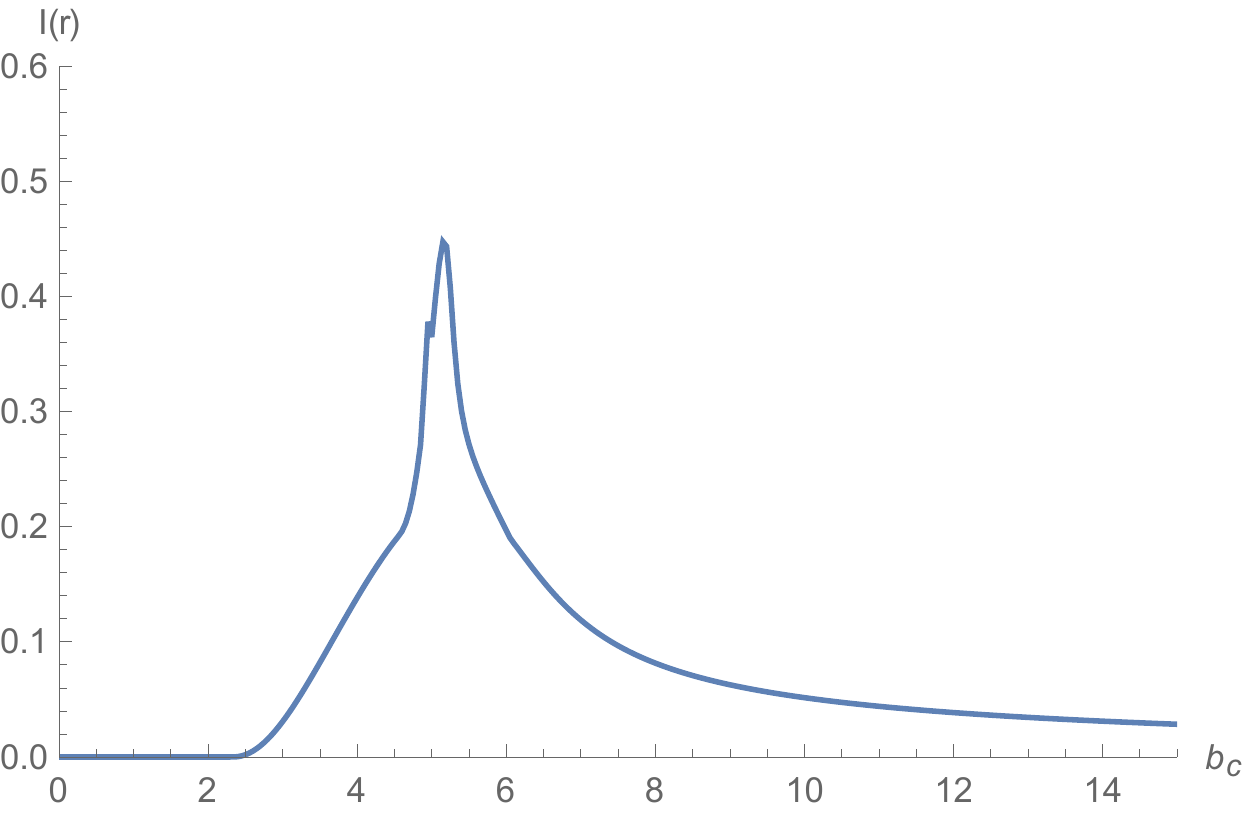}
\includegraphics[width=.2775\textwidth]{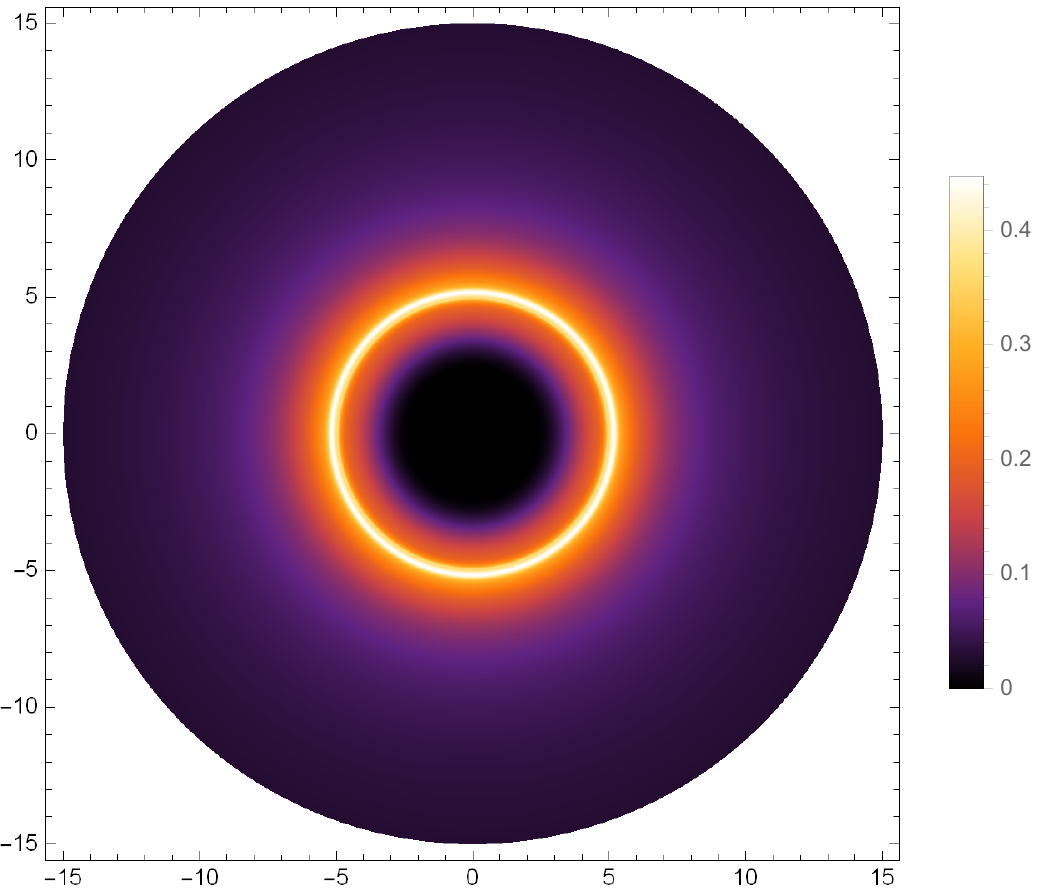}
\caption{\label{fig6}  Observational appearances of a geometrically and optically thin disk with different profiles near black hole. }
\end{figure}

After discussing the transfer function, then we will use Eqs.(\ref{eq182}) and (\ref{eq19}) to further study the observed specific intensity when some toy-model functions are given. As an example, the following typical toy-model functions are employed to discuss the observed specific intensity.
At first, we assume the emitted function $ I_{em}(r)$ is a decay function suppressed by the second power, which is
\begin{align}\label{eq20}
    I'_{{em}}(r) =\begin{cases}\left(\frac{1}{r-(r_{isco}-1)}\right)^2, &  r>r_{{isco}}  \\
    0, &r \leq r_{{isco}}
    \end{cases}
\end{align}
where $r_{isco}$ is the radius of the innermost stable circular orbit of KS black hole.
Secondly, we assume $I_{em}(r)$ is a decay function suppressed by the third power, which is
\begin{align}\label{eq21}
    I''_{em}(r) =\begin{cases}\left(\frac{1}{r-(r_{p}-1)}\right)^3, &  r> r_{p}   \\
    0, &r \leq r_{p}
    \end{cases}
\end{align}
with $r_{p}$ is the radius of the photon sphere.
Thirdly, we assume $I_{em}(r)$ is a moderate decay of emission, which is
\begin{align}\label{eq22}
    I'''_{em}(r) =\begin{cases} \frac{\frac{\pi }{2}-\tan ^{-1}(r-(r_{isco}-1))}{\frac{\pi }{2}+\tan ^{-1}(r_{p})}
    , &  r>r_+  \\
    0,  &r \leq r_+
    \end{cases}
\end{align}
and $r_{+}$ is the radius of the event horizon. By using those emitted functions, we have plotted Figures \ref{fig5} and \ref{fig6} to intuitively show the shadows and rings of KS black hole for different values of parameter $\omega$, where $\omega = 0.1$ for Figure \ref{fig5} and $\omega = 0.8$ for Figure \ref{fig6}.

In Figures \ref{fig5} and \ref{fig6}, the first, second and third row correspond to the emitted functions $I'_{em}(r)$, $I''_{em}(r)$ and $I'''_{em}(r)$, respectively. And, the left, middle and right column correspond to the emitted intensities, observed intensities and the corresponding appearances of KS black hole under various choices of the emitted function $I_{em}(r)$. In the first row of Figures \ref{fig5} and \ref{fig6}, it shows in the left column that the emission of thin disk will arrive at the peak at $r\simeq6M$ for $\omega =0.1$($r\simeq5.6M$ for $\omega=0.8$), and then abruptly decrease to zero at this point. This implies the emitted region is located at the position outside the photon sphere $r_p$. Because of the gravitational lensing, we can see from the middle column that, the peak and abruptly decay of the direct image of the thin disk occur at $ r\simeq6.9M $ for $\omega =0.1$($r\simeq6.4M$ for $\omega=0.8$). And, the observed lensing ring emission is presented as a thin ring with the location at $r\simeq5.4M $ for $\omega =0.1$($r\simeq5.2M$ for $\omega=0.8$), which contributes a very small part to the total observed intensities. Also, the photon ring\footnote{Where, this ring can be seen by magnifying the corresponding figure.} emission exhibits as a extremely narrow ring at $r\simeq5.2M $ for $\omega =0.1$($r\simeq4.9M$ for $\omega=0.8$), which makes a negligible contribution to the total observed intensities.
The third column as the appearance of KS black hole with the emitted functions $I'_{em}(r)$, presents the mainly observed intensity determined by direct emission, a smaller observed intensity of lensing ring and a hardly saw photon ring.

In the second row of Figures \ref{fig5} and \ref{fig6}, it implies that the emitted region has extended from $r<6M$ to the position near the photon sphere $r_p$. From the left column, we can see that the peak of emission occurs at $r\simeq3M$ for $\omega =0.1$($r\simeq2.7M$ for $\omega=0.8$). But, the observed intensity of direct emission has a peak at $r\simeq3.8M$ for $\omega =0.1$($r\simeq3.5M$ for $\omega=0.8$) from the middle column. And meanwhile, at the range $5.2M \thicksim 5.4M$($4.9M \thicksim 5.2M$ for $\omega=0.8$), the observed lensing ring and photon ring are superimposed that we can not distinguish them. In this case, similar to the first row, the total observed intensity is also dominated by the direct emission. The lensing ring contributes a small part, while the photon ring contributes a negligible part. Those features can be saw from the third column.

From the third row of Figures \ref{fig5} and \ref{fig6}, the left column describes the emitted region has extended to the position near the event horizon $r_+$, where $r\simeq2M$ for $\omega =0.1$($r\simeq1.6M$ for $\omega=0.8$). In this case, the observed intensity of the emission is very moderate by comparing with that of $I'_{em}(r)$ and $I''_{em}(r)$. And, it is true from the middle column that, the peak of the direct emission occurs at $r\simeq5.2M$ for $\omega =0.1$($r\simeq4.9M$ for $\omega=0.8$). Similarly, the lensing ring and photon ring are also superimposed. Although the contribution of lensing ring increases, the direct emission remains to dominate the total observed intensity. The photon ring also makes a negligible contribution.

By comparing Figure \ref{fig5} with Figure \ref{fig6}, we can conclude that the observational appearances of KS black hole exhibit some obvious different features for different values of $\omega$. It shows that the positions at which the peak of the observed intensity of direct emission occurs always decreases with the increase of parameter $\omega$, and the lensing ring and photon ring are all smaller and smaller when $\omega$ increased.
Also, for all three emitted functions, the maximum observed intensities of direct emission for $\omega =0.1$ are always higher than that for $\omega=0.8$.
However, the maximum observed intensity of the lensing ring is almost invariant with parameter $\omega$ for $I'_{em}(r)$, even if photon ring becomes brighter and brighter. Furthermore, since the lensing ring and photon ring superimposed for $I''_{em}(r)$ and $I'''_{em}(r)$, the maximum observed intensities of those rings always decreases for $I''_{em}(r)$, and increases a little for $I'''_{em}(r)$, when $\omega$ increased. Finally, it is worth noting that figure \ref{fig6} presents an interesting region occurred at the range $4.9M \thicksim 5.2M$ at which is superimposed by direct emission, lensing ring and pohton ring. Obviously, the observed intensity of this region is brighter than that of any other regions of direct emission, which is significantly different from that of the Schwarzschild black hole. Therefore, those results might be regarded as an effective characteristic for us to distinguish black holes in the HL gravity from the Schwarzschild black hole.

\section{Conclusions and discussions}\label{Sec4}
In deformed HL gravity, we in this paper have applied the ray-tracing method to carefully investigate the observational appearance of the KS black hole illuminated by the thin disk accretion.
First, we studied the effective potential and photon orbits of KS black hole by using the null geodesic.
Then, based on the definition of orbits $n = \phi/2\pi$, the trajectories of light rays emitted from the north pole direction are analysed.
Finally, when KS black hole was surrounded by the thin disk accretion which located in the rest frame of static worldlines,
we showed the first three transfer functions for different values of $\omega$, and the observed specific intensities of Direct emissions and rings by using three typical toy-model functions.
The result shows that, the event horizon $r_+$, the radius $r_p$ and impact parameter $b_{p}$ of photon sphere are all decreased with the increase of parameter $\omega$, and their value can be reduced to that obtained in the Schwarzschild black hole if one sets $\omega=0$.
Also, it turns out that the regions of the direct emission locates at $b_c < 5.00984M$ and $b_c>6.16591M$ for $\omega = 0.1$, and the lensing ring belongs to the range $5.00984M< b_c < 5.18382M$ and $5.22435M< b_c < 6.16591M$, while the range of photon ring is $5.18382M < b_c < 5.22435M $.
The related results for the case of $\omega = 0.8$ can be found in equation (\ref{Eq120}), and the corresponding trajectories of light ray have been presented in Figure \ref{fig3}.
More importantly, as black hole illuminated by the thin disk accretion, we present the size and observed luminosity of KS black hole shadows for different values of $\omega$ when emissions located at different positions in Figures \ref{fig4}, \ref{fig5} and \ref{fig6}. It shows that the lensed ring and photon ring always occur at the range $5.2M\sim5.4M$ for case $\omega=0.1$($4.9M\sim5.2M$ for case $\omega=0.8$), and they are distinguishable for $I'_{em}(r)$, but overlapped for $I''_{em}(r)$ and $I'''_{em}(r)$.
Further, our results show that the direct emissions always dominate the total observed intensity, while lensing rings as a thin ring make a very small contribution and photon ring as a extremely narrow ring make a negligible contribution, for all three toy-model functions.
Also, for different values of $\omega$, the features of observed intensity which are different from Schwarzschild black hole are presented finally.
Combined with those facts, we can conclude that the observational appearance pictured with some obvious different features in KS black hole, might be regarded as a characteristic for us to distinguish black holes in the HL gravity from the Schwarzschild black hole.
In addition, there may still exist the static and infalling cases of spherically symmetric accretion around the black hole in our universe. Therefore, it is interesting for us to further discuss the observational appearance of black holes surrounded by the static and infalling spherical accretions.

\vspace{10pt}

\noindent {\bf Acknowledgments}

\noindent
The authors would like to thank the anonymous reviewers for their helpful comments and suggestions, which helped to improve the quality of this paper.
This work is supported by the National Natural Science Foundation of China (Grant No. 11903025), and by the starting fund of China West Normal University (Grant No.18Q062).\\

\end{document}